\documentclass{aastex6}

\shorttitle{Resonant trapping in the galactic halo}
\shortauthors{Schuster, Fern\'andez-Trincado, \& Moreno}

\begin{document}

\title{Resonant Trapping of the Moving Groups G18-39 and G21-22 in the
Galactic Halo} 

\author{W. J. Schuster}
\affiliation{Instituto de Astronom\'\i{}a, Universidad Nacional
Aut\'onoma de M\'exico, Apdo. Postal 106, 22800 Ensenada, B.C.,
M\'exico.}

\author{J. G. Fern\'andez-Trincado}
\affiliation{Departamento de Astronom\'\i{}a, Casilla 160-C,
Universidad de Concepci\'on, Concepci\'on, Chile.}
\affiliation{Instituto de Astronom\'\i{}a y Ciencias Planetarias,
Universidad de Atacama, Copayapu 485, Copiap\'o, Chile.}
\affiliation{Institut Utinam, CNRS UMR 6213, Universit\'e
Bourgogne-Franche-Comt\'e, \\
OSU THETA Franche-Comt\'e, Observatoire de
Besan\c{c}on, BP 1615, 25010 Besan\c{c}on Cedex, France.}

\author{E. Moreno}
\affiliation{Instituto de Astronom\'\i{}a, Universidad Nacional
Aut\'onoma de M\'exico, Apdo. Postal 70-264, \\
Ciudad Universitaria CDMX 04510, M\'exico.}

\begin{abstract}

The 3D Galactic orbits of stars in the two groups G18-39 and
G21-22 pertaining to the Galactic halo have been computed in a Galactic
potential including a Galactic bar. The orbits have been related with
the orbital structure of resonant orbits on the Galactic plane created
by the bar component. We find that the majority of stars in both groups
are trapped mainly by two resonant families already studied in a
previous analysis. We show that the observed U--V velocity field of
the stars in both groups can be naturally explained as a result of
their trapping by these two resonant families, taking the angular
rotation speed of the bar approximately in the interval
45--60 $\mathrm{km\,s^{-1}\,kpc^{-1}}$. This analysis may help
to understand the identification of other known star groups as the
possible result of the interactions produced by resonances on stars
close to resonant families. For the two groups G18-39 and G21-22 we
conclude that the majority of their stars are members of the
supergroups of stars in the Galaxy trapped by two resonant families
generated by the Galactic bar.

\end{abstract}


\section{Introduction}
\label{intro}

\citet{1958MNRAS.118...65E} began a systematic study of moving groups
in our Galaxy using the convergent-point method to compile
stars of the young disk population in the solar neighborhood with
space motions similar to the Hyades cluster and the Sirius group, and
looked for the associated expected clumping of stars in the U--V
velocity plane \citep{1958MNRAS.118..560E}.
He found later that the Hyades and Sirius moving groups are more
extended localized groups which he called superclusters
\citep{1983AJ.....88..190E,1983AJ.....88..642E}, in which the star
members have parallel space motions and possibly come from the
disruption of compact stellar clusters. As the Galactic perturbations
on the clusters act over an increasing time, the star orbits distribute
along a tube of orbits; he called the resulting distribution of stars
a stellar group \citep{1996AJ....112.1595E}. For the old disk
population \citet{1991AJ....102.2028E} proposed that some studied
stellar groups are also associated with disrupted stellar clusters.
In general the identification of a stellar moving group formed by
this mechanism might not be an easy task, because the perturbed cluster
stars could follow orbits quite distinct from the original cluster 
orbit, i.e. the `tube' could be very wide, thus the expected clumping 
in the U--V plane would not be clear. The detailed chemical analysis
requiered in the identification of a disruption origin restricts the
connection of a moving group and a possible parent cluster; e.g.,
the results of \citet{2010AJ....139..636W,2015ApJ...808..103N}
respect to the possible association of the Kapteyn group with the
globular cluster $\omega$ Centauri. Considering a disruption on a
larger scale, a possible origin of moving groups in the Galactic halo
has been associated to the relics of satellite galaxies disrupted by
our own Galaxy \citep{1999Natur.402...53H,
2005MNRAS.359...93M,2014ApJ...787...31Z}.

Other proposed mechanisms for the origin of moving groups have
been considered in the literature. The effect of the spiral arms and
the Galactic bar has been studied by \citet{2009IAUS..254P...4A,
2009ApJ...700L..78A,2010HiA....15..192A,2011MNRAS.418.1423A,
2005AJ....130..576Q} using test-particle simulations to compare with
the observed distribution of disk stars in the U--V plane in the solar
neighborhood. For old moving groups in eccentric orbits, trapping of
stars by resonances due to spiral arms or a bar has been proposed by 
\citet{1998AJ....115.2384D}. A recent alternative mechanism has been
analyzed by \citet{2016MNRAS.458.4354O,2015MNRAS.447.3016O}, 
proposing that some moving groups in the solar neighborhood were born
by the desintegration of an old gaseous supercloud or by the capture
of stars due to a local giant interstellar cloud.

The present analysis is concerned in particular with the two moving
groups G18-39 and G21-22 in the Galactic halo, which were identified
by \citet{2012RMxAA..48..109S}. In that study the
suggested origin of these groups was related with the disruption of the
globular cluster $\omega$ Centauri, as inferred by theoretical
simulations made by \citet{2005MNRAS.359...93M}. Here we consider an
alternative explanation for the presence of these two groups in the
solar neighborhood, which is directly connected with resonant
trapping of 3D orbits by resonances on the Galactic plane generated
by the Galactic bar. We have analyzed this trapping mechanism in a
previous study \citep{2015MNRAS.451..705M}. In Section \ref{data} we
list the updated data for the stars in both groups G18-39 and G21-22.
The employed non-axisymmetric Galactic potential is presented in
Section \ref{modelo}, and the resonant trapping of stars in both groups
is analyzed in some detail in Section \ref{atrap}. The comparison
between the observed and resulting resonant U--V velocity distributions
is given in Section \ref{pUV}, and our conclusions are summarized in
Section \ref{concl}.

\section{Data for the two Moving Groups}
\label{data}

The kinematic parameters of the stars in the two groups G18-39 and
G21-22 are listed in \citet{2012RMxAA..48..109S}. Some of these
parameters have been updated using recent published data. The radial
velocities have been selected from the
literature using always the most precise and accurate values available,
generally with errors less than $\pm 1.0\, \mathrm{km\,s^{-1}}$, such
as those of \citet{2002AJ....124.1144L}, \citet{2017arXiv170709322A},
\citet{2010A&A...511L..10N}, and \citet{2004A&A...424..727P}.
\citet{2006AstL...32..759G} has given radial velocities with errors
over a range $\pm 0.2$ to $\pm 4.8$ $\mathrm{km\,s^{-1}}$, and
for the more southern stars the RAVE survey \citep{2017AJ....153...75K,
2011AJ....141..187S,2008AJ....136..421Z} errors over $\pm 1.0$
to $\pm 2.8$ $\mathrm{km\,s^{-1}}$. Nine stars, which were included
in the catalogue employed in our previous publication related with
moving groups \citep{2015MNRAS.451..705M}, still retain their older
radial velocities with errors from $\approx \pm 5$ to
$\approx \pm 7$ $\mathrm{km\,s^{-1}}$. The median error of the radial
velicities used here is $\pm 0.28$ $\mathrm{km\,s^{-1}}$.

Table~\ref{tabla1} lists the parameters employed in our computations.
The star identification is that given in \citet{2012RMxAA..48..109S}.
A convenient reference number is listed in the second column, and in the
following columns appear the positions, radial velocity, proper motions,
uncertainties in the radial velocity and proper motions, and distance
from the Sun with uncertainty.

\tabletypesize{\scriptsize}
\tablewidth{0pt}
\tablecolumns{12}
\begin{deluxetable*}{ccrrrrrcrrrr}
\tablecaption{Data for the two moving groups \label{tabla1}}
\tablehead{
\colhead{ID} & \colhead{Number} & \colhead{$\alpha$} & \colhead{$\delta$} & \colhead{$V_\mathrm{rad}$} & \colhead{$\mu_{\alpha}$} & \colhead{$\mu_{\delta}$} & \colhead{$\sigma V_{\mathrm{rad}}$} & \colhead{$\sigma \mu_{\alpha}$} & \colhead{$\sigma \mu_{\delta}$} & \colhead{$D$} & \colhead{$\sigma D$}  \\
\colhead{}  & \colhead{}  & \colhead{(degrees)} & \colhead{(degrees)} & \colhead{($\mathrm{km\,sec^{-1}}$)} & \colhead{($\mathrm{mas\,yr^{-1}}$)} & \colhead{($\mathrm{mas\,yr^{-1}}$)} & \colhead{($\mathrm{km\,sec^{-1}}$)} & \colhead{($\mathrm{mas\,yr^{-1}}$)} & \colhead{($\mathrm{mas\,yr^{-1}}$)} & \colhead{(pc)} & \colhead{(pc)} 
}
\startdata
\cutinhead{GROUP G18-39}
G31-26 & 1 & 2.0818292 & $-5.2485278$ & $-217.05$ & 353.000000 & $-123.000000$ & 0.3 & 5.000000 & 5.000000 & 200.3 & 20.0  \\
HD3567 & 2 & 9.6332007 & $-8.3115530$ & $-48.20$ & 20.664299 & $-546.521926$ & 0.2 & 0.089203 & 0.062938 & 115.3 & 4.0 \\
G70-33 & 3 & 15.9760000 & $-3.8539444$ & $-83.42$ & 342.000000 & $-69.000000$ & 0.3 & 5.000000 & 5.000000 & 246.3 & 24.6 \\
G34-45 & 4 & 25.7032083 & 22.6173056 & $-277.68$ & 87.380000 & $-304.720000$ & 0.2 & 10.000000 & 10.000000 & 133.7 & 16.6 \\
LP709-053 & 5 & 33.0099844 & $-14.0082561$ & $-7.00$ & 263.849965 & $-62.525949$ & 7.0 & 2.173218 & 1.119031 & 263.9 & 26.1 \\
G36-47 & 6 & 44.3338936 & 26.2805580 & 88.93 & 260.765338 & $-226.759889$ & 0.2 & 1.606972 & 0.747114 & 224.5 & 24.8 \\
G86-39 & 7 & 80.7978333 & 33.1841944 & 214.30 & 412.865000 & $-720.029000$ & 0.3 & 10.000000 & 10.000000 & 81.1 & 8.1 \\
G98-53 & 8 & 93.4577354 & 33.4158976 & 144.49 & 22.658701 & $-325.715595$ & 0.2 & 5.754996 & 6.037369 & 177.7 & 7.1 \\
$-48:2445$ & 9 & 100.3614907 & $-48.2198975$ & 319.60 & 64.566503 & 217.261808 & 7.0 & 1.083189 & 0.596698 & 186.2 & 9.6 \\
G87-13 & 10 & 103.7345667 & 35.5162722 & 206.53 & 53.500000 & $-241.700000$ & 0.2 & 3.900000 & 3.800000 & 256.3 & 25.6 \\
G89-014 & 11 & 110.6316879 & 8.8191592 & $-36.02$ & 151.047078 & $-269.304977$ & 1.2 & 0.157074 & 0.082924 & 220.5 & 12.0 \\
LP490-061 & 12 & 160.9967490 & 12.8011075 & 213.00 & $-149.289457$ & $-97.293485$ & 7.0 & 2.422196 & 1.275121 & 418.9 & 129.3 \\
G120-15 & 13 & 166.5854875 & 31.2140639 & 131.53 & $-202.680000$ & $-126.650000$ & 0.2 & 10.000000 & 10.000000 & 318.0 & 31.8 \\
G10-03 & 14 & 167.5109792 & $-2.7905361$ & 202.20 & 141.000000 & $-468.
000000$ & 0.8 & 5.000000 & 5.000000 & 111.9 & 13.9 \\
$-24:9840$ & 15 & 174.0112167 & $-24.8728417$ & 164.05 & $-21.200000$ & $-215.700000$ & 1.0 & 3.100000 & 2.900000 & 309.3 & 30.9 \\
HD101063 & 16 & 174.4168530 & $-28.8514150$ & 183.40 & $-313.626643$ & $-15.760882$ & 0.3 & 0.076786 & 0.051859 & 249.6 & 17.5 \\
G176-53 & 17 & 176.6464773 & 50.8818562 & 64.40 & $-870.070000$ & $-543.760000$ & 0.2 & 1.080000 & 0.890000 & 76.0 & 7.6 \\
G66-30 & 18 & 222.5325179 & 0.8408762 & $-115.08$ & $-283.780000$ & $-104.710000$ & 0.1 & 2.870000 & 2.470000 & 233.1 & 23.3 \\
G15-24 & 19 & 232.6723763 & 8.3940935 & $-88.51$ & $-391.277812$ & $-116.995738$ & 0.2 & 0.340198 & 0.191280 & 159.3 & 6.8 \\
LP636-003 & 20 & 311.2551750 & $-1.6820528$ & $-57.00$ & $-38.300000$ & $-186.100000$ & 7.0 & 5.500000 & 5.500000 & 339.8 & 33.9 \\
G187-30 & 21 & 317.8235693 & 33.5249133 & $-340.60$ & 502.936813 & 159.151139 & 0.6 & 0.189435 & 0.313993 & 99.4 & 2.7 \\
$-03:5166$ & 22 & 319.3293250 & $-2.7486250$ & $-122.00$ & $-103.400000
$ & $-248.000000$ & 7.0 & 2.500000 & 2.300000 & 260.4 & 26.0 \\
G18-39 & 23 & 334.6533014 & 8.4453971 & $-234.77$ & 283.316895 & $-102.1
84230$ & 0.2 & 0.212514 & 0.149986 & 154.6 & 5.8 \\
LP877-025 & 24 & 343.3748958 & $-23.9131306$ & 117.78 & $-24.700000$ & $-400.500000$ & 9.5 & 2.300000 & 2.100000 & 195.4 & 19.5 \\
G157-85 & 25 & 354.1300417 & $-8.4312222$ & $-82.0$ & 254.000000 & $-144.000000$ & 7.0 & 5.000000 & 5.000000 & 258.7 & 25.9 \\
\cutinhead{GROUP G21-22}
G02-38 & 1 & 21.7298294 & 12.0056807 & $-171.16$ & $-10.988957$ & $-359.179492$ & 0.3 & 3.316039 & 0.712774 & 185.6 & 17.2 \\
G82-42 & 2 & 70.5000208 & $-4.2794694$ & $-8.62$ & 2.500000 & $-347.600000$ & 0.3 & 4.600000 & 4.200000 & 218.2 & 21.8 \\
G108-48 & 3 & 105.4036445 & 6.4080035 & $-87.77$ & 3.130000 & $-674.110000$ & 0.3 & 3.200000 & 2.390000 & 133.4 & 13.3 \\
HIP36878 & 4 & 113.7224627 & $-10.3881404$ & 89.00 & 424.679519 & $-482.042040$ & 4.8 & 0.135706 & 0.118253 & 106.2 & 2.8 \\
G116-45 & 5 & 146.1589536 & 38.6096750 & $-33.89$ & 210.121664 & $-238.618536$ & 0.2 & 1.179395 & 1.077341 & 303.3 & 74.3 \\
G161-73 & 6 & 146.4082615 & $-4.6756279$ & 120.92 & 150.156734 & $-251.937567$ & 0.2 & 2.187828 & 1.692420 & 279.1 & 48.4 \\
G146-56 & 7 & 159.9603750 & 42.1514444 & $-104.40$ & 81.710000 & $-212.
860000$ & 0.3 & 10.000000 & 10.000000 & 310.0 & 31.0 \\
G119-64 & 8 & 168.2003636 & 35.7267285 & $-196.25$ & 71.001036 & $-510.780372$ & 0.2 & 0.122860 & 0.092605 & 139.5 & 4.5 \\
G139-16 & 9 & 257.4474167 & 8.0737500 & 40.77 & $-100.000000$ & $-410.000000$ & 0.2 & 14.000000 & 7.000000 & 197.2 & 19.7 \\
LP808-022 & 10 & 267.6524583 & $-16.9841111$ & 200.00 & $-60.000000$ & $-216.000000$ & 7.0 & 5.000000 & 5.000000 & 315.4 & 31.5 \\
G154-25 & 11 & 267.6541667 & $-16.9866667$ & 200.00 & $-60.500000$ & $-216.400000$ & 7.0 & 5.500000 & 5.500000 & 313.3 & 31.3 \\
G21-22 & 12 & 279.7904792 & 0.1206528 & 59.45 & $-168.900000$ & $-446.200000$ & 0.2 & 1.500000 & 1.500000 & 158.9 & 15.9 \\
\enddata
\end{deluxetable*}

\section{The Galactic Model}
\label{modelo}

The Galactic orbits of the stars in both groups G18-39 and G21-22
were computed in a Galactic potential including axisymmetric and
non-axisymmetric components. The base for this potential is the
axisymmetric model of \citet{1991RMxAA..22..255A}. All the mass in
the spherical bulge component in this model has been employed to
construct a Galactic prolate bar with a similar distribution of mass
(i.e., the surfaces of equal density are concentric spheroids with
the same eccentricity), whose potential is given in
\citet{2004ApJ...609..144P} and has a density law and geometry which
approximates a model of \citet{1998ApJ...492..495F} of
\textit{COBE}/DIRBE observations of the Galactic center. After this
transformation, the whole potential, including the axisymmetric
components, was rescaled to the Sun's
galactocentric distance $R_0=8.3\, \mathrm{kpc}$ and LSR (Local
Standard of Rest) velocity
$\Theta_0=239\, \mathrm{km\,s^{-1}}$ \citep{2011AN....332..461B}.
The employed prolate bar, with a mass of
1.6$\times 10^{10} M_{\odot}$, is an approximation for the more
detailed model of the inner Galactic region given by
\citet{2017MNRAS.465.1621P}, which consists of a boxy/peanut bulge
aligned with a long bar, both components giving a total mass around
1.88$\times 10^{10} M_{\odot}$.

The present orientation of the major axis of the Galactic bar and its
angular rotation speed was revised in our earlier study of resonant
trapping in the Galactic halo \citep{2015MNRAS.451..705M}. There the
values 20$^{\circ}$ and 55 $\mathrm{km\,s^{-1}\,kpc^{-1}}$ were
considered for these two parameters. Here we extend their possible
range of values inferred by recent analysis of bar parameters based on
the study of the Hercules stream
\citep{2018MNRAS.474...95H,2017ApJ...840L...2P}, and the model of 
\citet{2017MNRAS.465.1621P} mentioned above, which propose values
of the orientation angle up to around 30$^{\circ}$ and angular speeds
down to around 40 $\mathrm{km\,s^{-1}\,kpc^{-1}}$. We include these
extensions in our computations. Table~\ref{tabla2} summarizes the
values of the parameters in the employed non-axisymmetric Galactic
model, including the used Solar motion \citep{2010MNRAS.403.1829S},
with $U$ negative towards the Galactic center.

\tabletypesize{\scriptsize}
\tablewidth{0pt}
\tablecolumns{2}
\begin{deluxetable*}{lc}
\tablecaption{Parameters Employed in the Non-Axisymmetric Galactic
Potential \label{tabla2}}
\tablehead{\\
\multicolumn{2}{c}{Position and Velocity of the LSR and Solar Velocity}}
\startdata
$R_0$    & 8.3 $\pm 0.23\, \mathrm{kpc}\,\,\,\, $\tablenotemark{(a)}  \\
${\Theta}_0$  & 239 $\pm 7.0\, \mathrm{km\,s^{-1}}\,\,\,\,$\tablenotemark{(a)}  \\
$(U,V,W)_{\odot}$ & ($-11.10$,12.24,7.25) $\pm (1.20,2.10,0.60) \mathrm{km\,s^{-1}}\,\,\,\,$\tablenotemark{(a),(b)}  \\
\cutinhead{Properties of the Galactic Prolate Bar}
Law Density   & $\rho(a) \propto sech^2(a)\,\,\,\,$\tablenotemark{(c),(d)}; $\,a$=major semiaxis of prolate similar surfaces  \\
Mass (scaled Bulge Mass in  &      \\
\citet{1991RMxAA..22..255A} Galactic Model)  & 1.6$\times 10^{10} M_{\odot}\,\,\,\,$\tablenotemark{(e)}  \\
Present position of major axis      & 20$^{\circ}$,25$^{\circ}$,30$^{\circ}\,\,\,\,$\tablenotemark{(f)}  \\
Angular velocity $\Omega_b$   & 40,45,50,55,60 $\mathrm{km\,s^{-1}\,kpc^{-1}}\,\,\,\,$\tablenotemark{(f),(g),(h)} \\
\enddata
\tablenotetext{(a)}{$\,\,\,\,$\citet{2011AN....332..461B}}
\tablenotetext{(b)}{$\,\,\,\,$\citet{2010MNRAS.403.1829S}}
\tablenotetext{(c)}{$\,\,\,\,$\citet{1998ApJ...492..495F}}
\tablenotetext{(d)}{$\,\,\,\,$\citet{2004ApJ...609..144P}}
\tablenotetext{(e)}{$\,\,\,\,$\citet{1991RMxAA..22..255A}}
\tablenotetext{(f)}{$\,\,\,\,$\citet{2018MNRAS.474...95H}}
\tablenotetext{(g)}{$\,\,\,\,$\citet{2017ApJ...840L...2P}}
\tablenotetext{(h)}{$\,\,\,\,$\citet{2017MNRAS.465.1621P}}
\end{deluxetable*}

\section{Resonant Trapping of the two Moving Groups}
\label{atrap}

In this section we consider the first part of our analysis of the
two groups G18-39 and G21-22, focussing in the possible trapping of
the stars in these groups by resonances on the Galactic plane generated
by the Galactic bar, which may help to understand their identification
as star groups.

\subsection{The Characteristic Energy--J plane}
\label{EJp}

In our earlier study of resonant trapping in the Galactic halo
\citep[hereafter Paper I]{2015MNRAS.451..705M}, the orbits of 1642
halo and disk stars in the solar neighborhood were computed in the
non-inertial reference frame where the bar component of the potential
is at rest. In that study we employed the boxy bar given in
\citet{2004ApJ...609..144P}. Every orbit was computed backward in time
during 10 Gyr and a characteristic energy--$J$ diagram was found to be
useful in the orbital analysis. $J$ is the orbital Jacobi constant
per unit mass, and the characteristic energy was defined as
($E_{\rm min}+E_{\rm max}$)/2, with $E_{\rm min}$, $E_{\rm max}$ the
minimum and maximum values of the non-constant orbital energy per unit
mass with respect to the \textit{inertial} Galactic frame. In this
diagram, which employes in general 3D orbits, some accumulations of
points appear in certain regions, which were found to be related with
corresponding positions in this diagram of families of periodic orbits
on the Galactic plane, i.e., 2D orbits, generated by the bar. Thus, in
these regions, 3D orbits may be trapped by resonant 2D orbits on the
Galactic plane.

In the present study the orbits of the stars in both groups were also
computed backward in time during 10 Gyr for each value of the angular
rotation speed of the Galactic bar, $\Omega_b$, listed in
Table~\ref{tabla2}, and with their kinematic results characteristic
energy--J diagrams were plotted in each case. To have a background of
comparison, the orbits of the 1642 stars employed in Paper I were again
represented in these diagrams (in fact, the members of the two groups
G18-39 and G21-22 are included in the 1642 stars sample). The results
given in this and the following sections correspond to an initial
orientation of 20$^{\circ}$ for the major axis of the Galactic bar;
the values 25$^{\circ}$ and 30$^{\circ}$ gave similar orbits in both
groups.

\begin{figure*}
\gridline{\fig{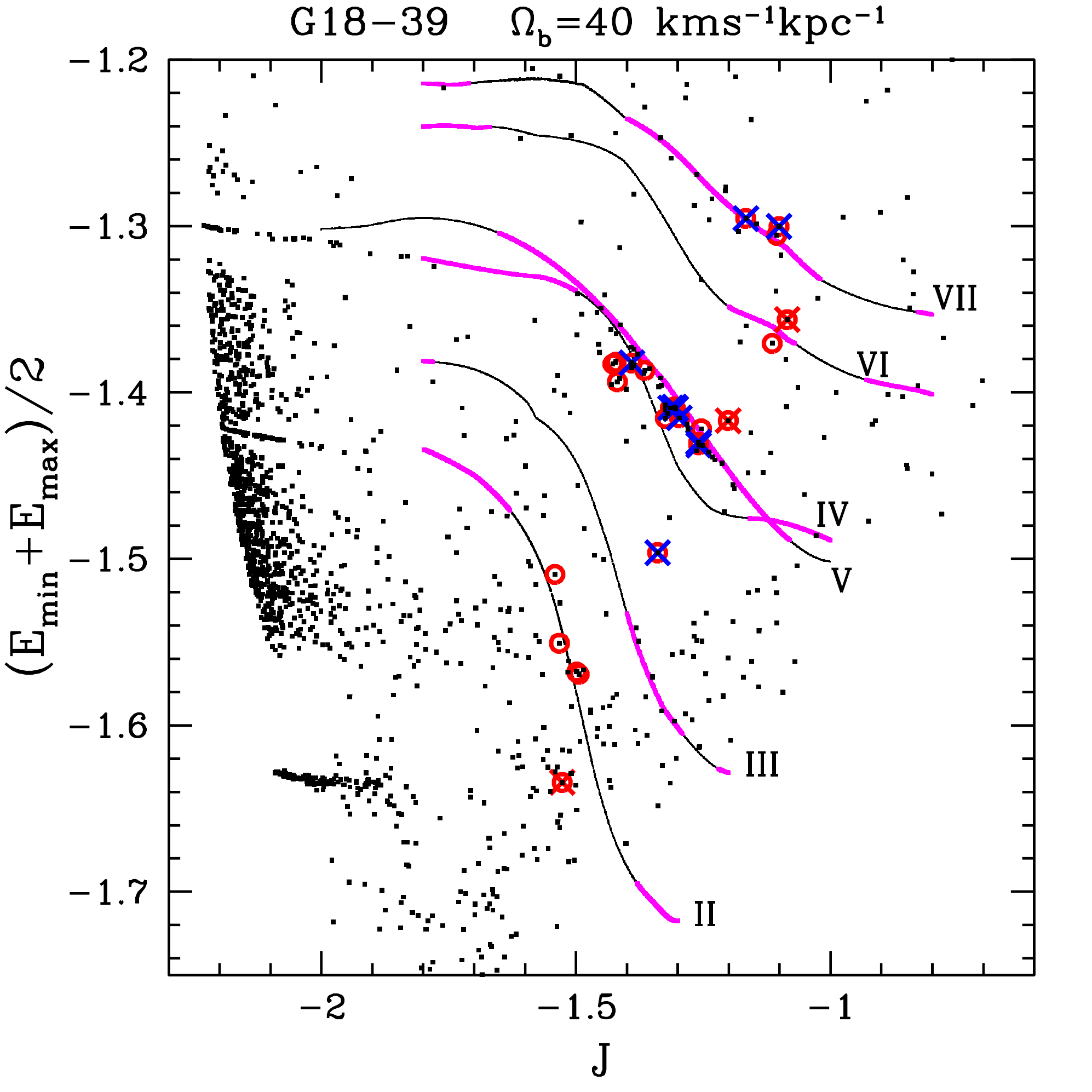}{0.33\textwidth}{(a)}
          \fig{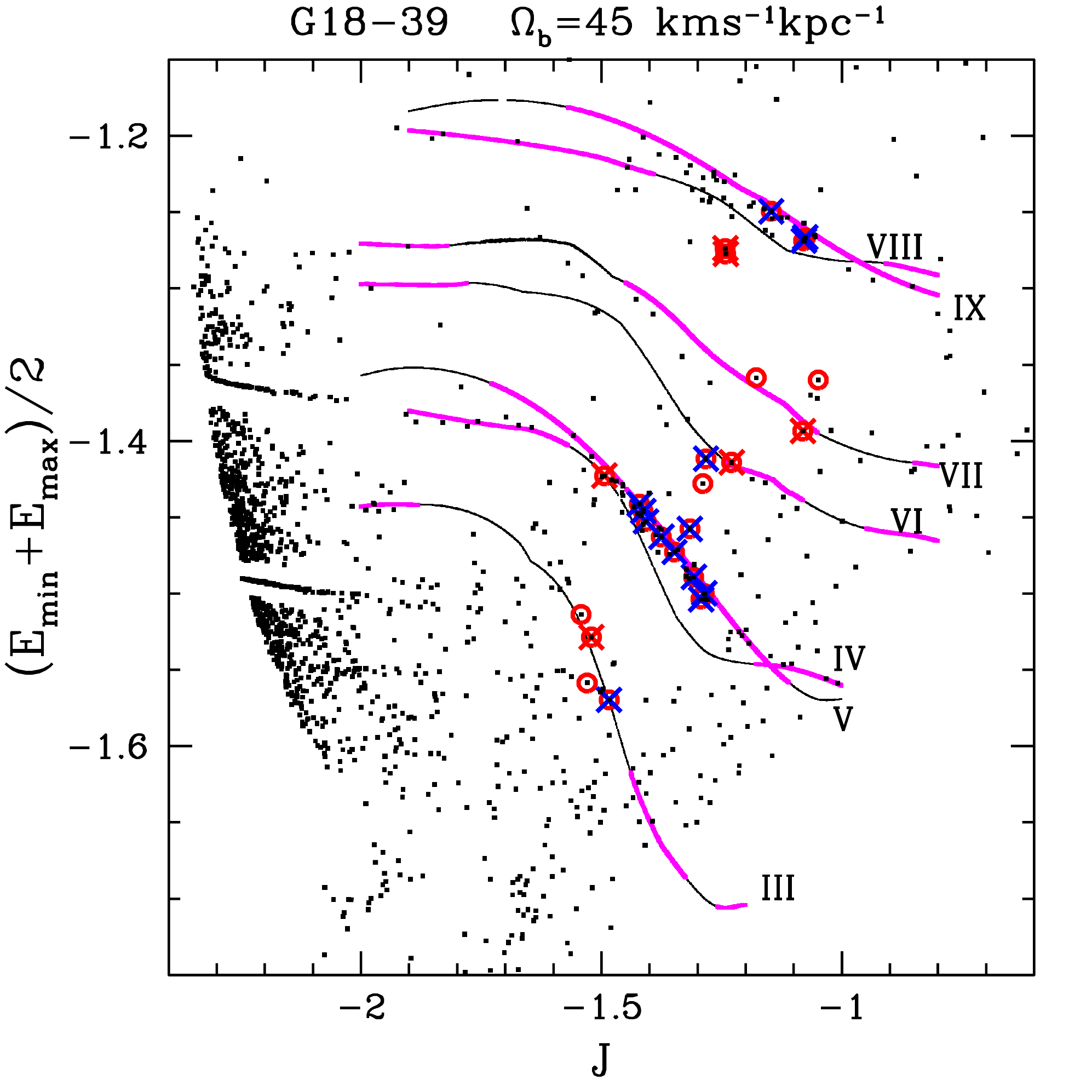}{0.33\textwidth}{(b)}
          \fig{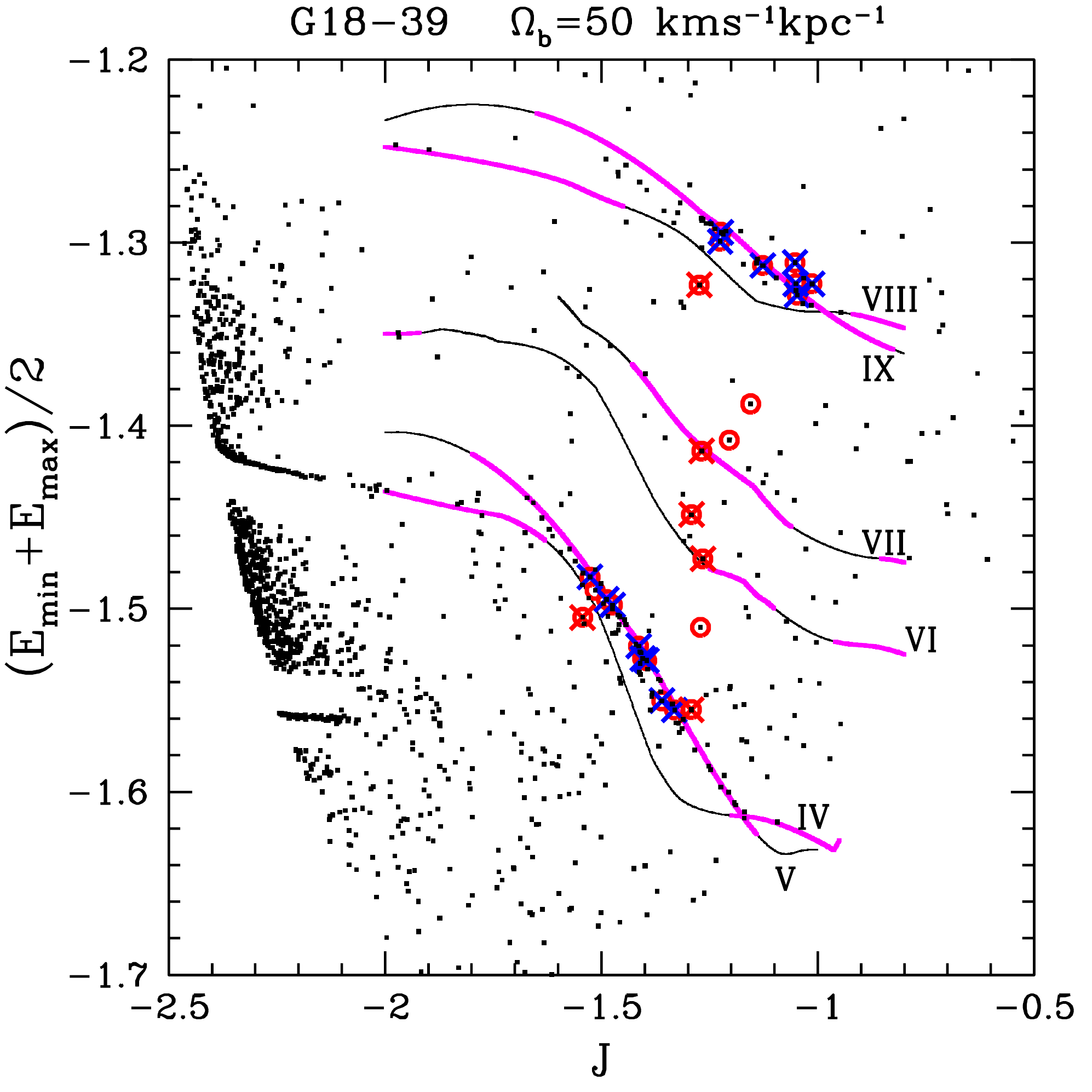}{0.33\textwidth}{(c)}}
\gridline{\fig{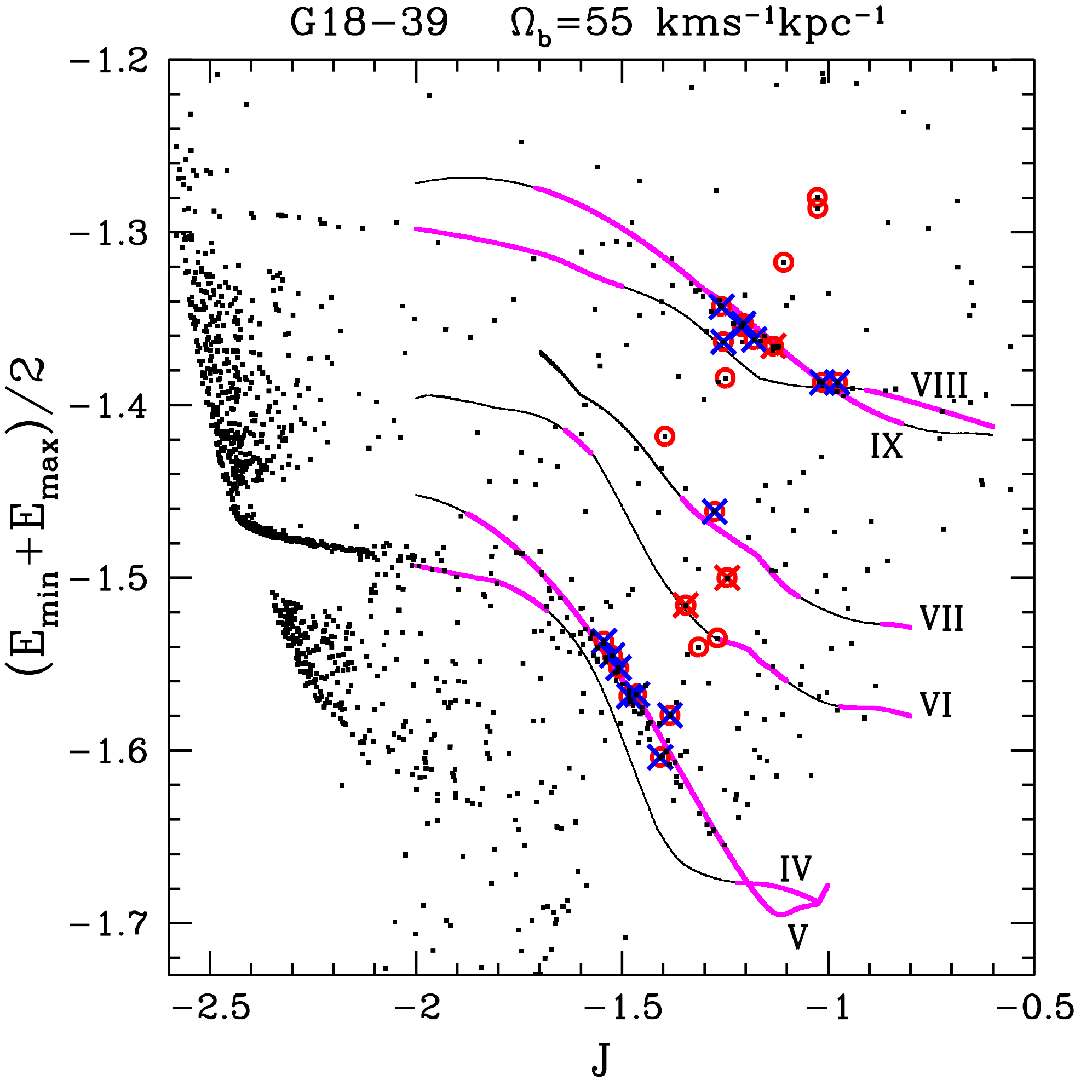}{0.33\textwidth}{(d)}
          \fig{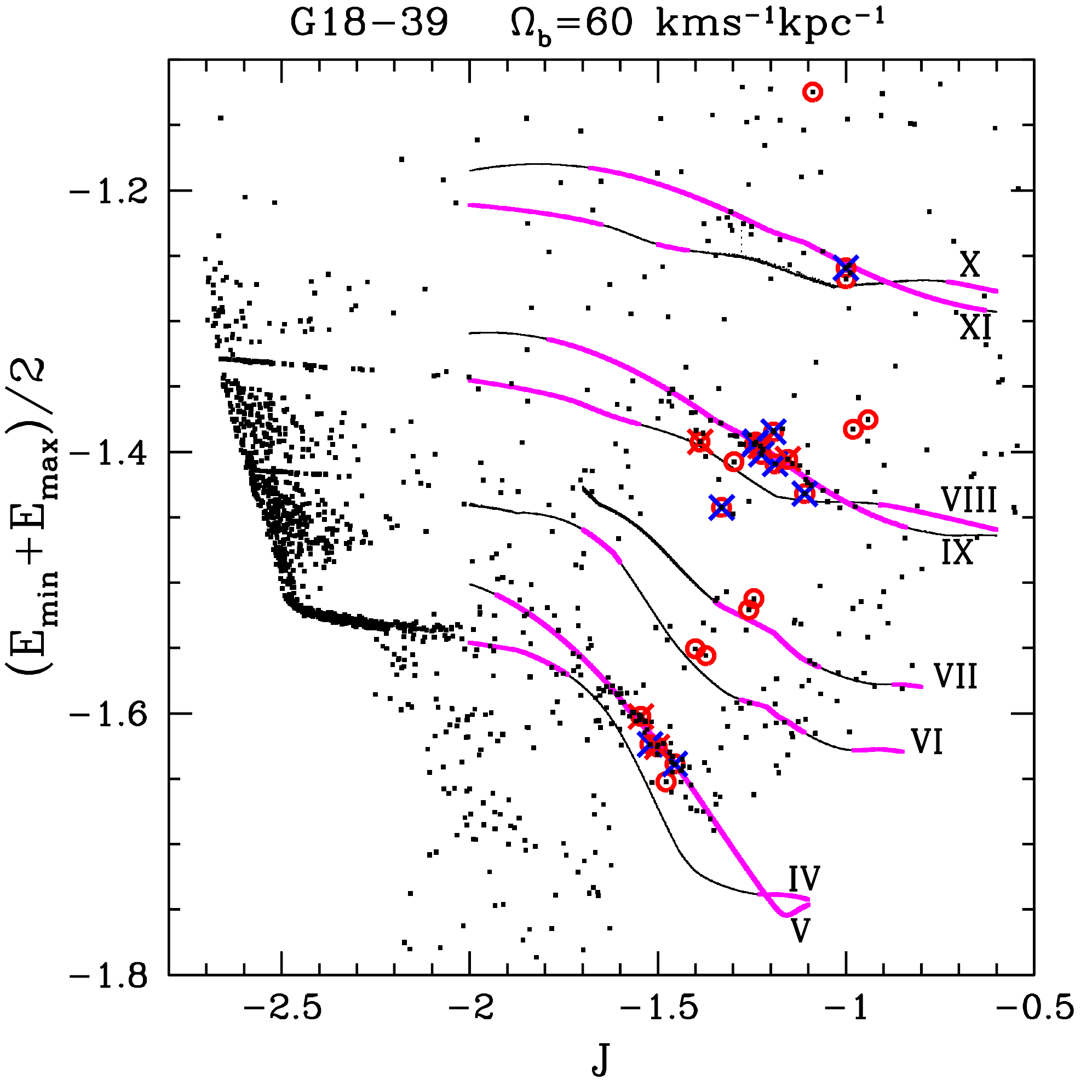}{0.33\textwidth}{(e)}}
\caption{Characteristic energy--J diagram for the group G18-39 using
$\Omega_b$=40,45,50,55,60 $\mathrm{km\,s^{-1}\,kpc^{-1}}$. The units in
($E_{\rm min}+E_{\rm max}$)/2 and $J$ are
 10$^5$ $\mathrm{km^2\,s^{-2}}$.
The small black points correspond to the 1642 stars sample. The curves
are some of the main 2D resonant families generated by the Galactic
bar. The magenta and black parts of each curve are stable and unstable
points along the family, respectively. Each curve is numbered as in
Paper I. The red-circled points belong to the group. Of these points,
those with a blue cross represent orbits trapped by resonances during
a time interval greater than 5 Gyr, the points with a red cross are
orbits trapped during less than 5 Gyr; points without a cross represent
orbits not trapped by resonances. See main text. \label{fig1}}
\end{figure*}

\begin{figure*}
\gridline{\fig{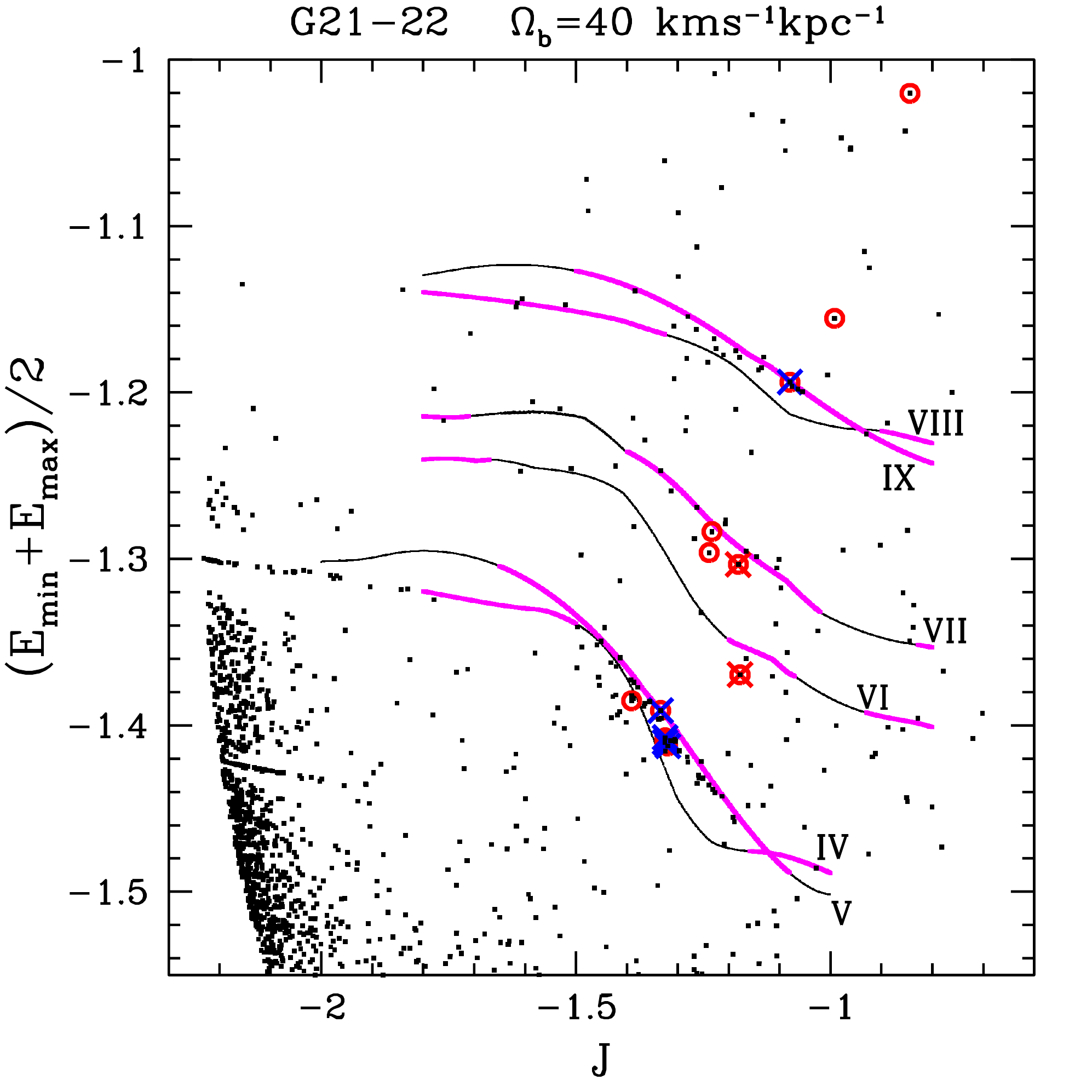}{0.33\textwidth}{(a)}
          \fig{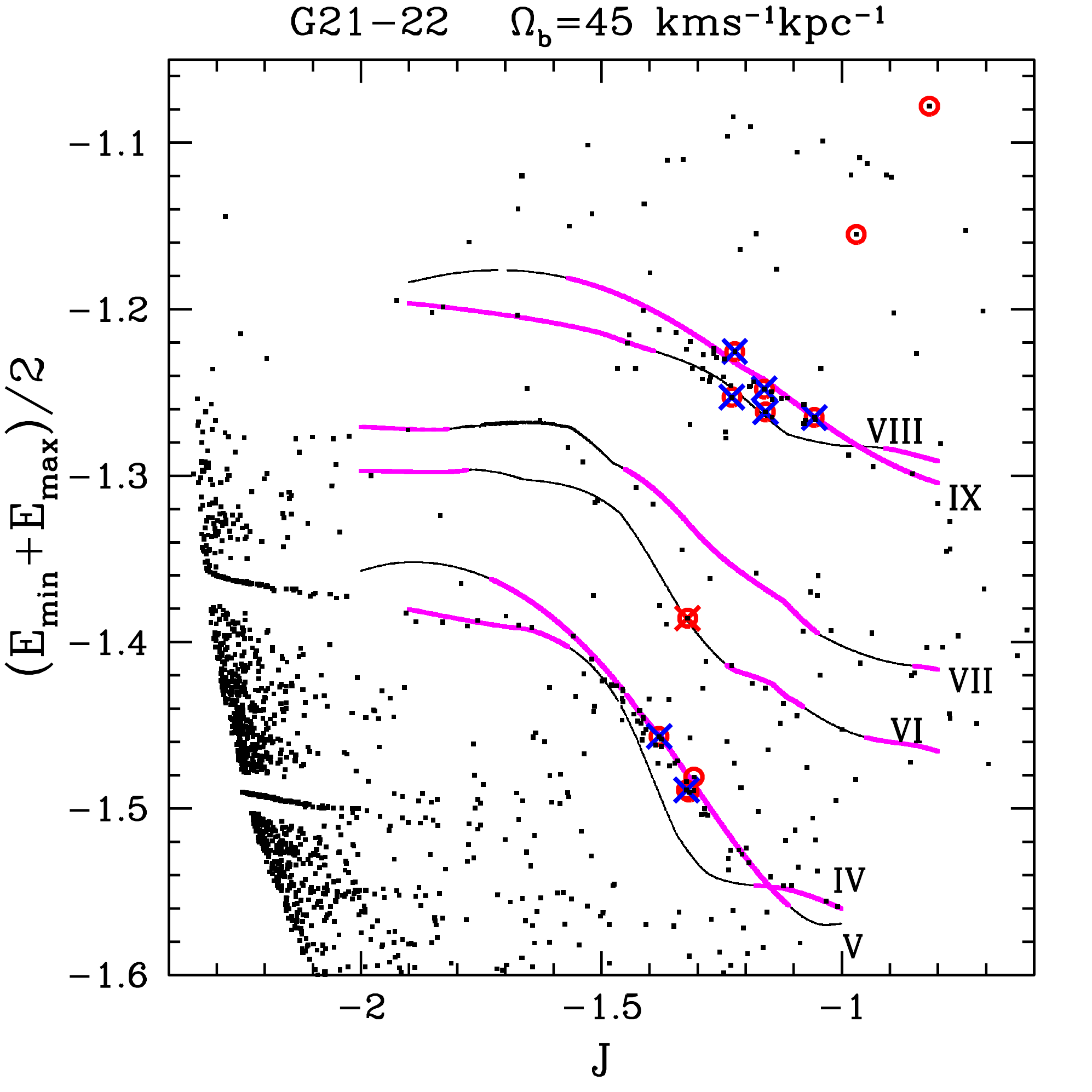}{0.33\textwidth}{(b)}
          \fig{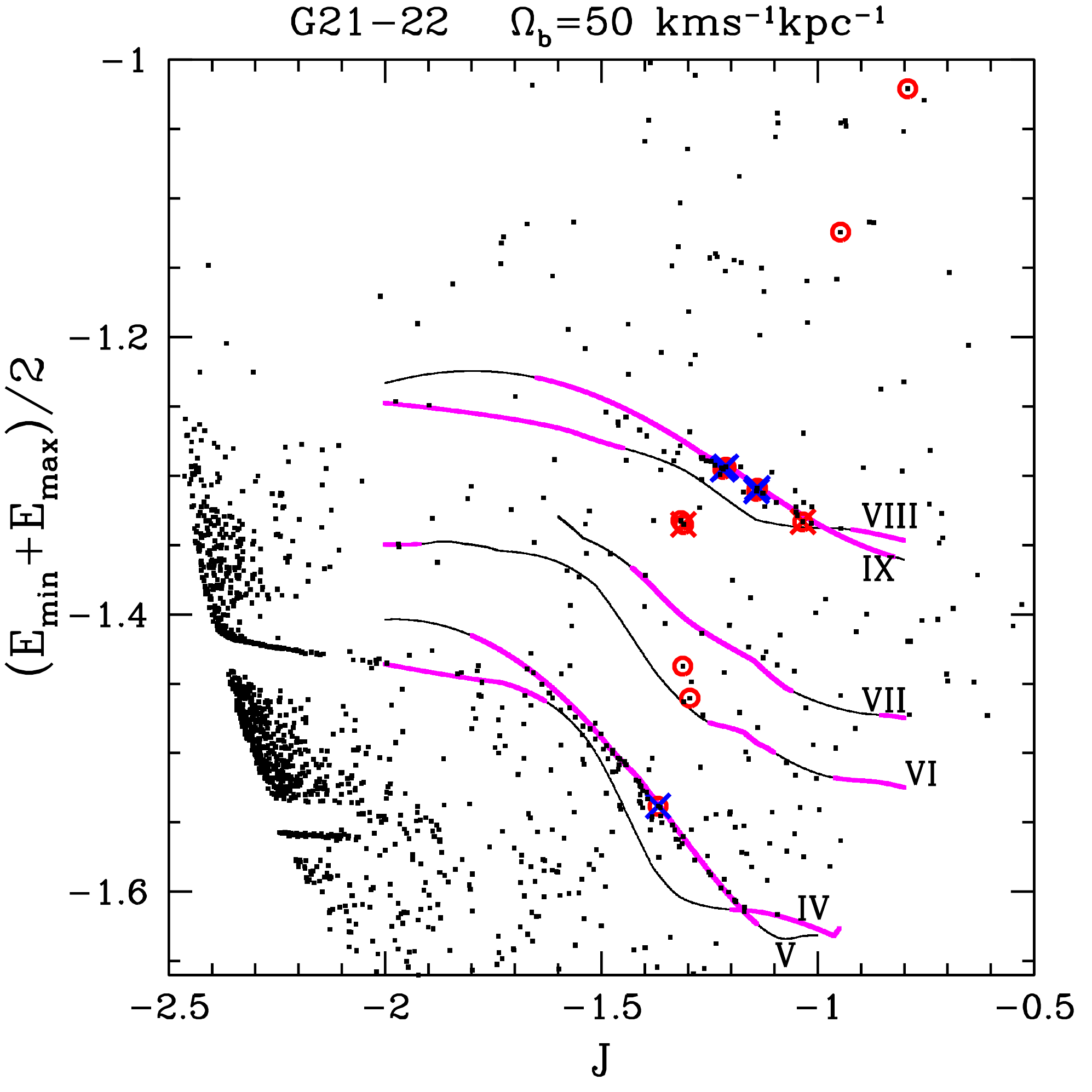}{0.33\textwidth}{(c)}}
\gridline{\fig{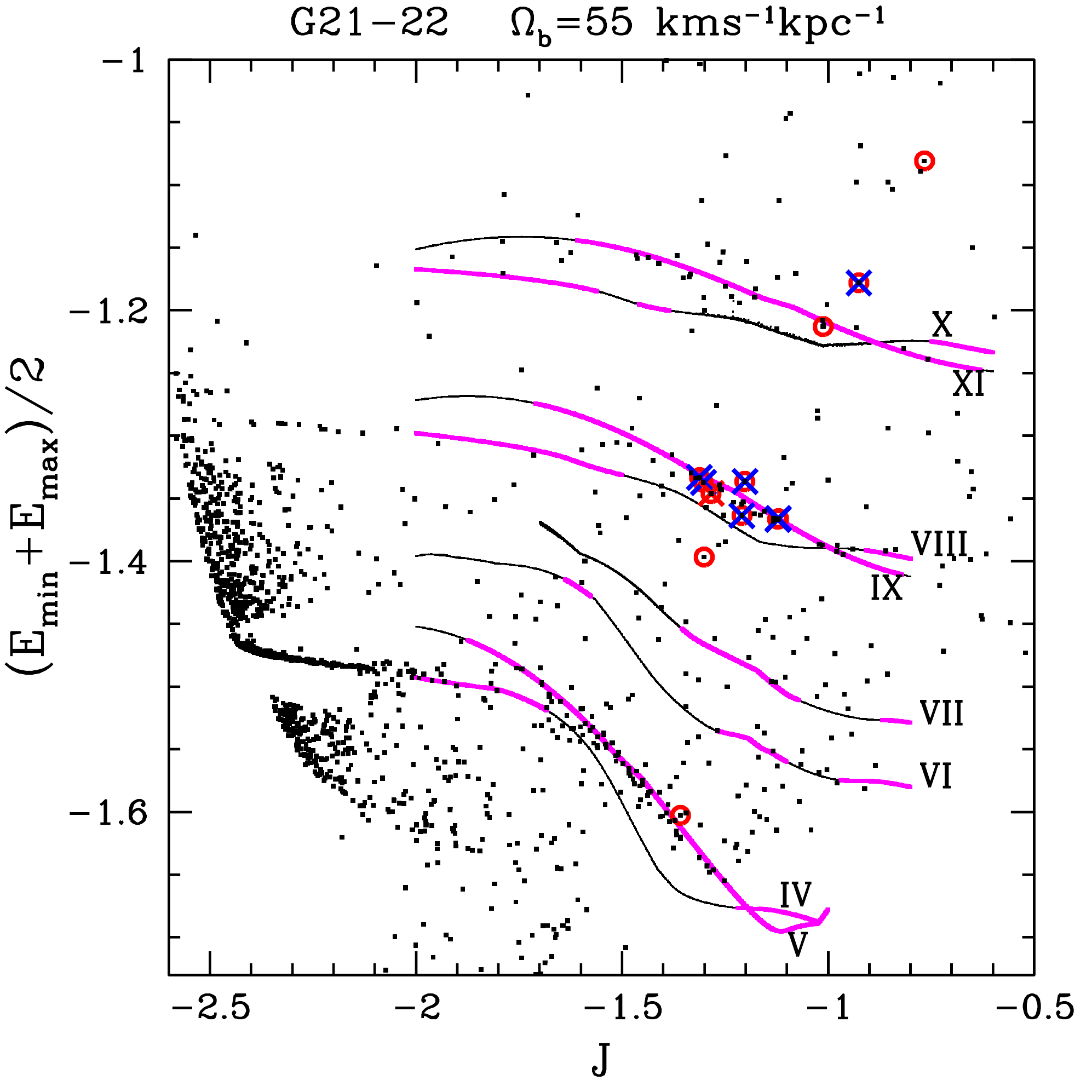}{0.33\textwidth}{(d)}
          \fig{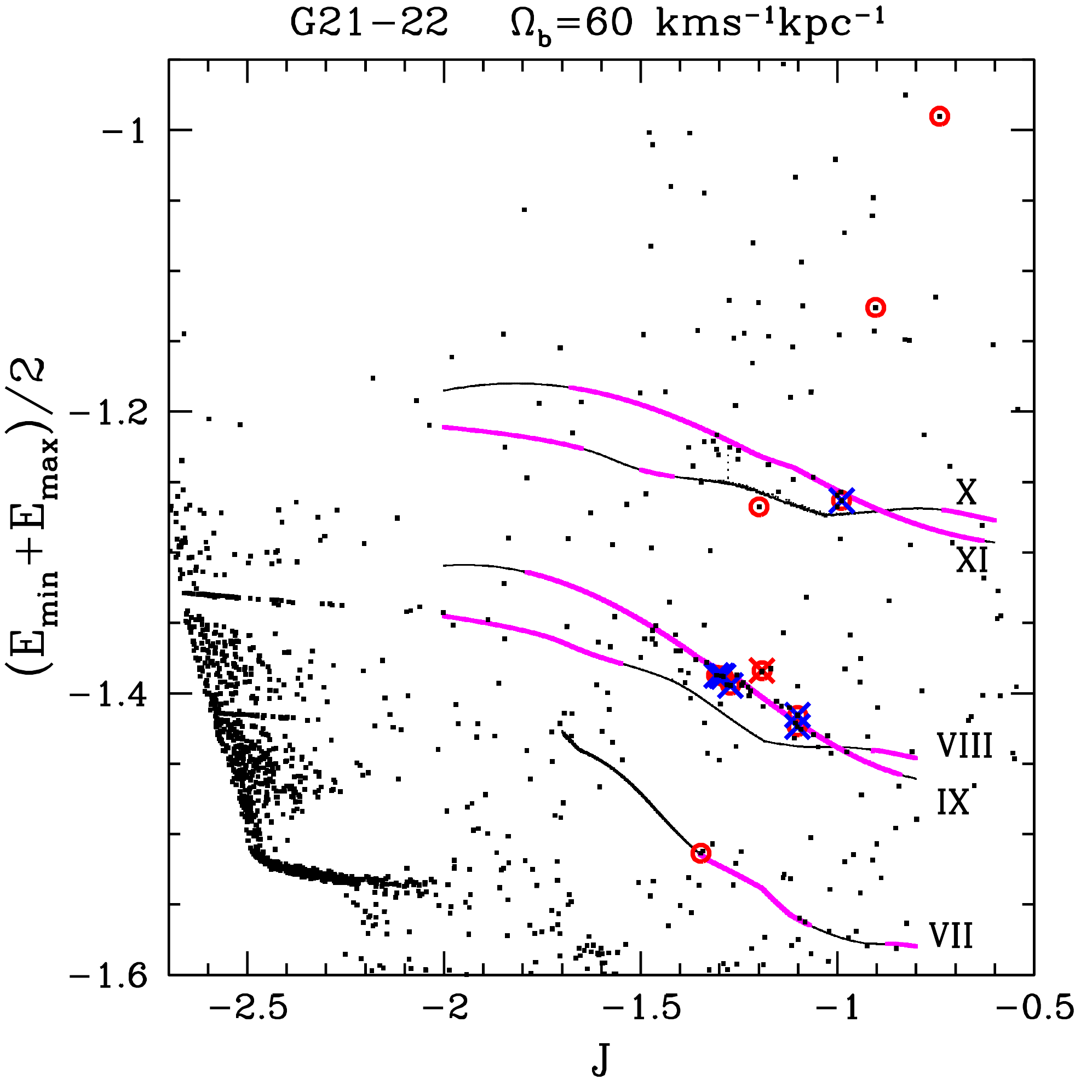}{0.33\textwidth}{(e)}}
\caption{As in Figure~\ref{fig1}, here the corresponding diagrams
for the group G21-22. \label{fig2}}
\end{figure*}

In Figures~\ref{fig1} and \ref{fig2} we show the characteristic
energy--J diagrams, which
include the results obtained for the two groups under different values
of the rotation speed $\Omega_b$.
The units employed for ($E_{\rm min}+E_{\rm max}$)/2 and $J$ are
10$^5$ $\mathrm{km^2\,s^{-2}}$. The small black points are part of the
1642 stars sample. In these diagrams it is useful to show the loci of
points or curves corresponding to some of the main 2D resonant families
generated by the Galactic bar, as well as the stable and unstable
parts of each curve. Some of these curves are shown in
Figures~\ref{fig1} and \ref{fig2}.
As in Paper I, the curves were computed using Poincar\'e diagrams
combined with a Newton-Raphson scheme, and the type of stability
computed with the method given by \citet{1965AnAp...28..992H}; see also
\citet{GC2002}.
The curves are numbered with notation in Paper I. The magenta and black
parts of each curve correspond respectively to stable and unstable
points along the family. Only parts of these curves are shown; they
can be extended to the left and right in all figures.
The red-circled points in these figures, some of which have a blue or
red cross, correspond in each case to one of the two analyzed groups,
and are commented in the following subsections. Some important features
can be noticed in the characteristic energy--J diagrams shown in
Figures~\ref{fig1} and \ref{fig2}: 
(a) for each value of $\Omega_b$ the points of both groups,
and all the background points, distribute in different ways, (b) the
resonant family curves move downwards as $\Omega_b$ is increased and
new families appear at the top, (c) the majority of points in a given
group tend to accumulate around some resonant families.

\subsection{The two Moving Groups in the Characteristic Energy--J plane}
\label{EJpl}

To represent the points of a given group in the characteristic
energy--J diagrams, the corresponding orbits were analyzed to see if
they were trapped or not by resonances. A first indication is obtained
with a visual inspection of the orbits. During the total computed
time of 10 Gyr each 3D star orbit in a group was plotted projected on
the Galactic plane, as viewed in the \textit{non-inertial} reference
frame where the bar is at rest. This projection was compared during
several subintervals of time with different orbital forms of 2D
resonant families due to the bar, which are already given in figures
presented in Paper I. As in general the projection will not coincide
with the closed curve of a 2D resonant orbit, the visual criterion to
establish when a 3D orbit was trapped by a given resonant family
consisted in the comparison of the projected orbit with expected
associated tube orbits on the Galactic plane around \textit{stable}
points of this family.

\begin{figure}[t!]
\plotone{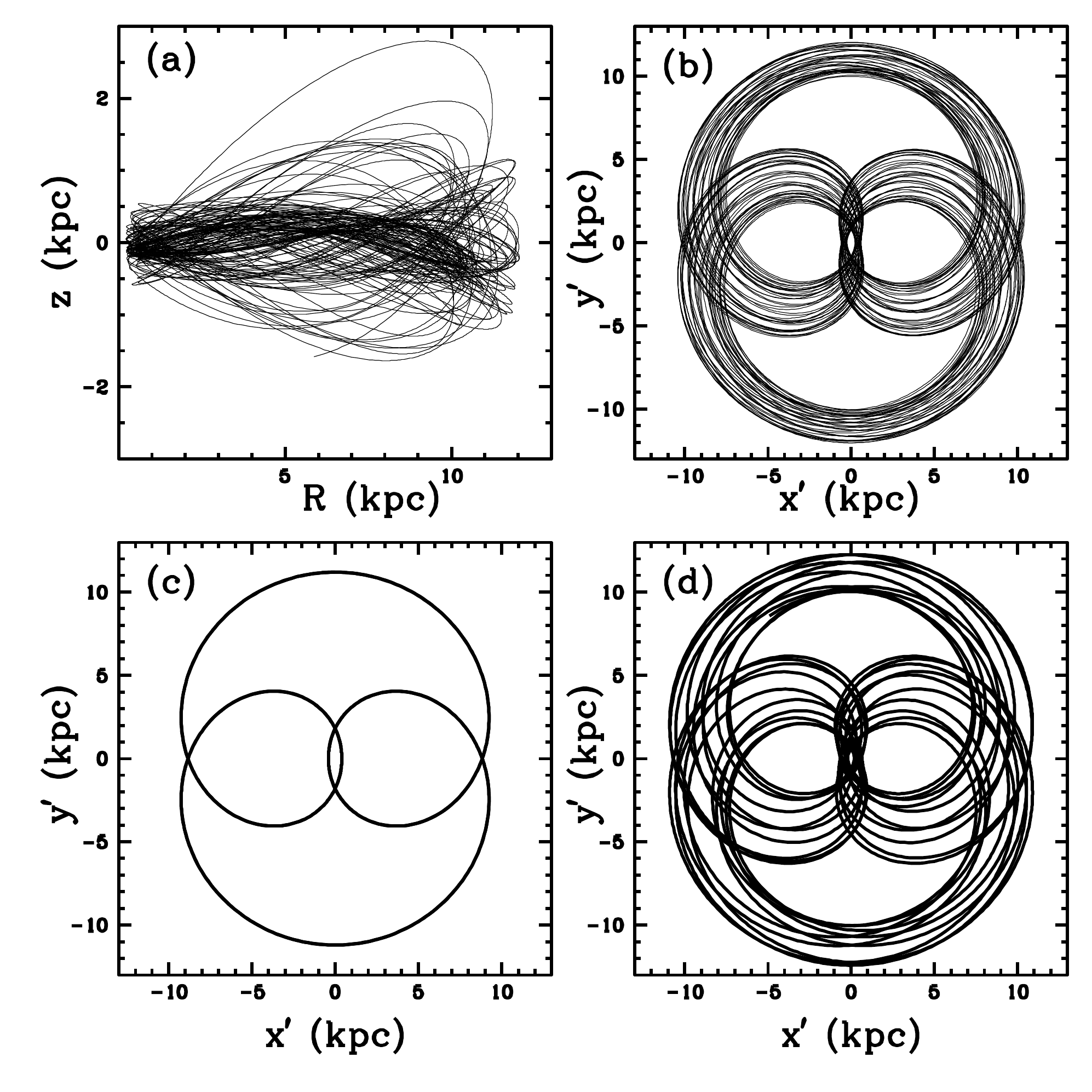}
\caption{An example of a trapped orbit; for star number 11 in the group
G18-39, with $\Omega_b = 50 ~\mathrm{km\,s^{-1}\,kpc^{-1}}$.
The meridional orbit is shown in frame (a) and the projected orbit
on the Galactic plane $x^{\prime},y^{\prime}$ (with the
$x^{\prime}$-axis pointing along the major axis of the bar) in the
non-inertial reference frame of the bar is shown in (b). This projected
orbit is associated with the resonant family V; see figure 14 in
Paper I. Frame (c) shows the \textit{stable} resonant orbit in family V
with the same value of the Jacobi constant $J$ as in the 3D orbit of
the star. Frame (d) shows a tube orbit on the Galactic plane around
this resonant orbit. A visual comparison of frames (b) and (d) shows
that the 3D orbit is trapped by family V. \label{fig3}}
\end{figure}

Figure~\ref{fig3} shows an example of an orbit trapped by a resonance
according to a visual inspection. The 3D orbit corresponds to star
number 11 in Table~\ref{tabla1} for the group G18-39. The value of
$\Omega_b$ is 50 $\mathrm{km\,s^{-1}\,kpc^{-1}}$. Frame (a) shows
the meridional orbit, and frame (b) shows the orbit projected on the
Galactic plane as viewed in the non-inertial reference frame of the
bar. The $x^{\prime}$-axis points along the major axis of the bar.
The orbit is clearly trapped by a resonance on the Galactic plane.
The trapping family is the one numbered as V in the corresponding
characteristic energy--J diagram shown in frame (c) of
Figure~\ref{fig1}, as can
be inferred by the comparison of the projected orbit with the forms of
resonant 2D orbits in this family given in Paper I (see figure 14 in
that paper). For the same value of the Jacobi constant $J$ in the 3D
orbit, frame (c) in Figure~\ref{fig3} shows the corresponding 2D
\textit{stable} resonant orbit in family V, and frame (d) shows a tube
orbit on the Galactic plane around this resonant orbit which resembles
the projected orbit in frame (b). Thus, we conclude that the orbit is
trapped by family V.

We considered a second, complementary and more quantitative, criterion
to see if an orbit is trapped or not by resonances. As was done
in the case shown in Figure~\ref{fig3}, for the resulting value
of $J$ in the 3D orbit of a star in a group the corresponding 2D
\textit{stable} resonant orbit of the inferred trapping family was
localized. A series of tube orbits on the Galactic plane were computed
around this resonant orbit and were represented in a Poincar\'e
diagram. In particular, to construct this diagram we localized the
points where the orbits cross the $y^{\prime}$-axis. This diagram
corresponds to 2D orbits on the Galactic plane; thus, the 3D orbit
of the star is not directly represented in this diagram. We chose the
natural extension of localizing the points where the 3D orbit crosses
the $x^{\prime}=0$ plane (at different $z^{\prime}$ positions with
respect to the Galactic plane), and with the corresponding values of
the position $y^{\prime}$ and velocity $v_{y^{\prime}}$ this orbit was
plotted in the Poincar\'e diagram. In this diagram the points of the 3D
orbit will not necessarily lie on a given invariant curve, as the 2D
tube orbits do, but we expect that if the 3D orbit is trapped, its
points will distribute among the curves representing the tube orbits.

\begin{figure}[t!]
\plotone{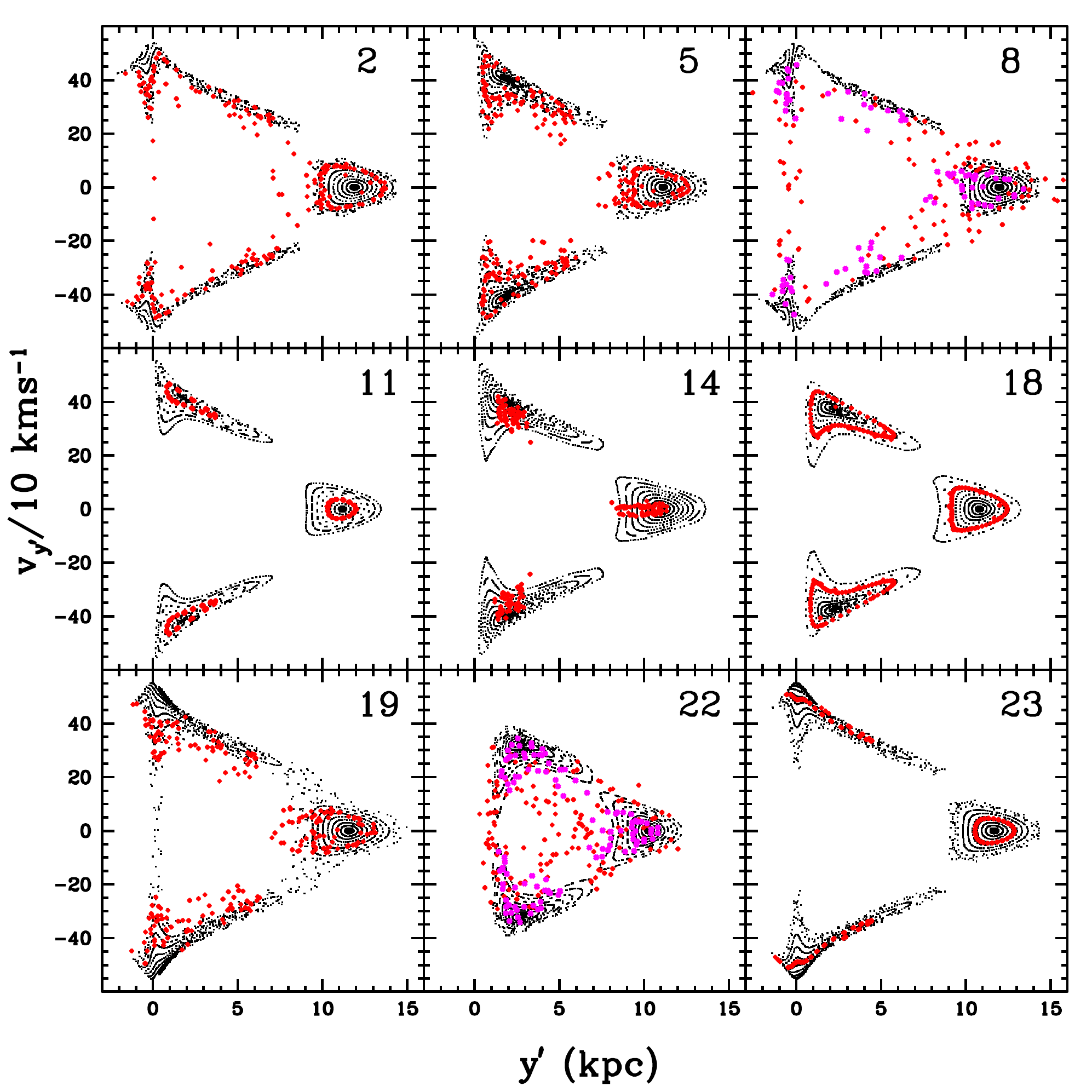}
\caption{Examples of extended Poincar\'e diagrams for stars in the
group G18-39 with $\Omega_b = 50 ~\mathrm{km\,s^{-1}\,kpc^{-1}}$.
The number in each frame is the number of star in the group. The black
points show some tube orbits around the stable periodic orbit in the
trapping family (family V in this figure), with the same value of $J$
as the 3D orbit. The red points correspond to the 3D orbit
in the group. The two frames with points in magenta color are cases in
which the orbit is approximately trapped during a time interval less
than 10 Gyr, shown by these magenta points. \label{fig4}}
\end{figure}

\begin{figure}[t!]
\plotone{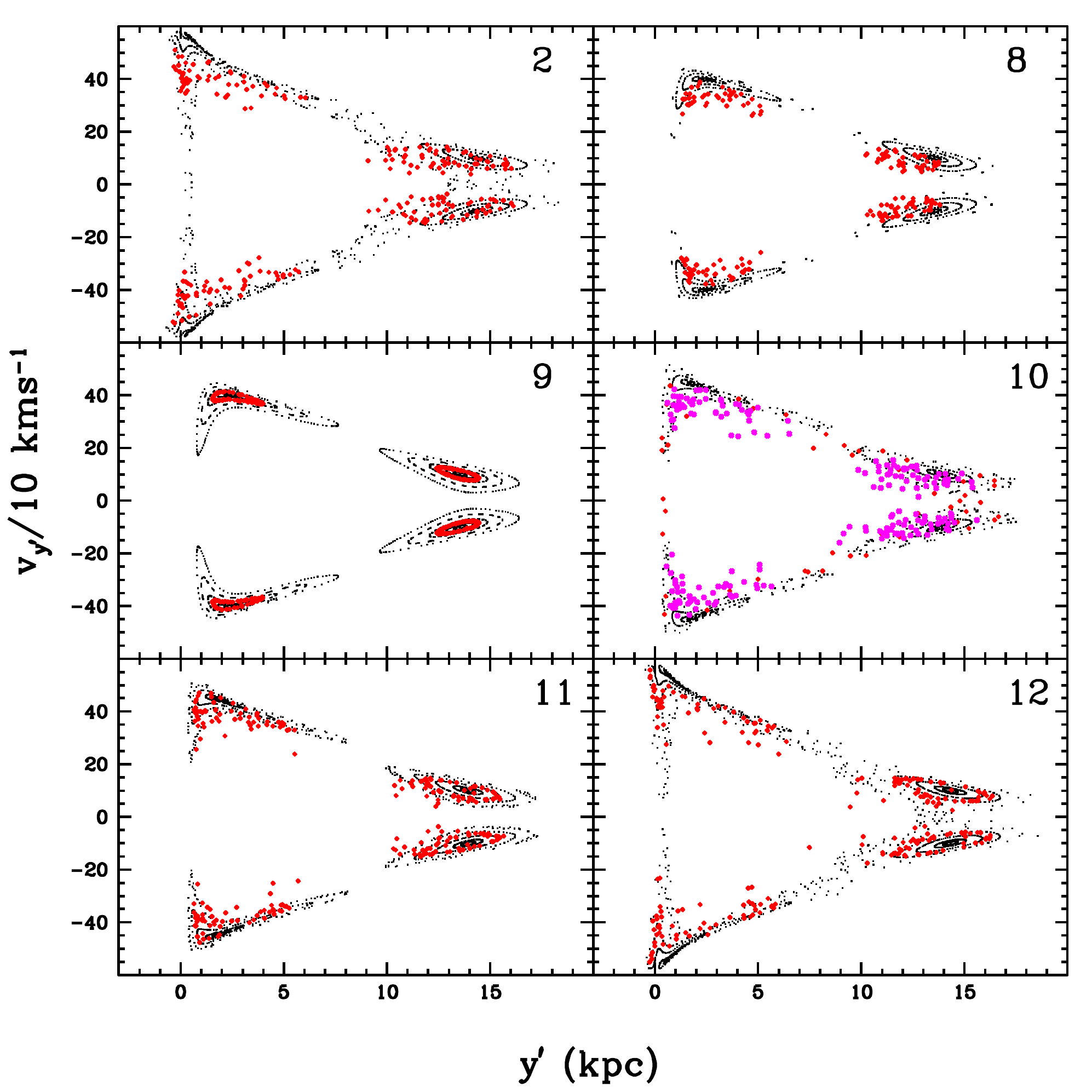}
\caption{As in Figure~\ref{fig4}, here for the group G21-22 with
$\Omega_b = 55 ~\mathrm{km\,s^{-1}\,kpc^{-1}}$. The trapping family in
this figure is family IX. \label{fig5}}
\end{figure}

\begin{figure}[t!]
\plotone{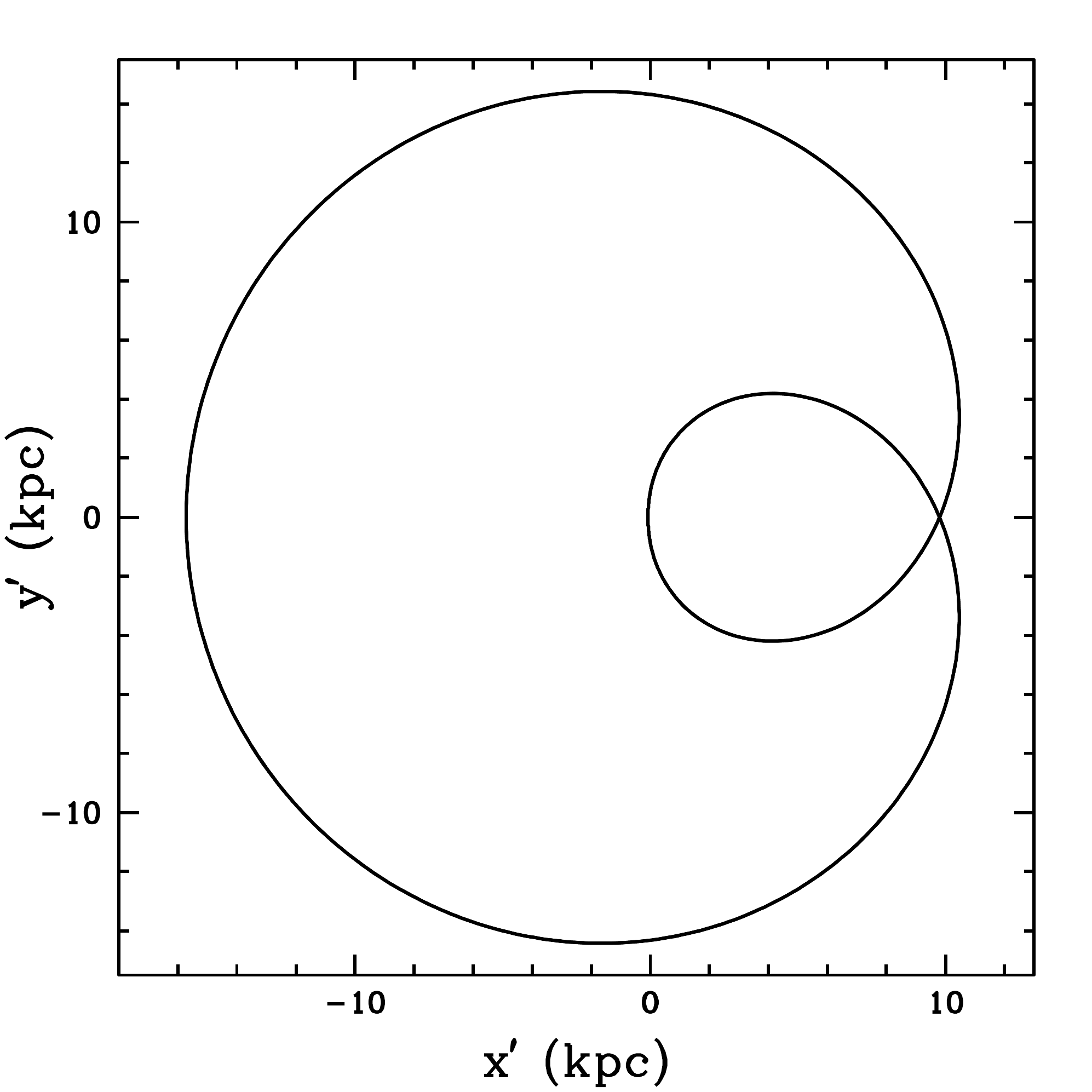}
\caption{An example of a periodic orbit in family IX. \label{fig6}}
\end{figure}

To illustrate this last procedure, Figures~\ref{fig4} and \ref{fig5}
 show some
Poincar\'e diagrams associated to orbits of some stars in groups G18-39
and G21-22 which by the visual criterion, and by the results presented
in these diagrams, have been considered as trapped by resonant orbits
on the Galactic plane. In Figure~\ref{fig4} nine examples are shown for
the group G18-39 with $\Omega_b = 50 ~\mathrm{km\,s^{-1}\,kpc^{-1}}$,
and Figure~\ref{fig5} shows six examples for the group G21-22 with
$\Omega_b = 55 ~\mathrm{km\,s^{-1}\,kpc^{-1}}$. The number in each
frame is the number of star in the group; see Table~\ref{tabla1}.
The trapping family in Figure~\ref{fig4} is family V, and in
Figure~\ref{fig5} is family IX. The orbital form of periodic orbits
in these families can be seen in figures 14 and 18 in Paper I;
Figure~\ref{fig6} gives an example of a periodic orbit in this
family IX.
The black points in each frame of Figures~\ref{fig4} and \ref{fig5} 
correspond to some tube orbits around the stable periodic
orbit. The red points correspond to the 3D orbit in the group, which
approximately distribute among the curves of the tube orbits. In
particular, the frame with number 11 in Figure~\ref{fig4} gives the
result for the star in Figure~\ref{fig3}. The three examples in
Figures~\ref{fig4} and \ref{fig5} in which
some points appear with a magenta color indicate cases where the orbit
is trapped only during some interval of time, less than the 10 Gyr
considered in the orbit; the magenta points correspond to that limited
interval. For all considered values of $\Omega_b$ we found that some
of the 3D orbits could be trapped only in some intervals of time,
as in their projection on the plane $x^{\prime},y^{\prime}$ they were
oscilating inside and outside a tube orbits region. Thus, if the
projected orbit was approximately inside the tube region in a
certain interval of time, this time was considered as a contribution
for the total trapping time of the corresponding resonant trapping
family. In the majority of cases the total trapping time was equal to
the total computed orbital time of 10 Gyr. In Section \ref{orbnatr}
some more detail is given on the behavior of an orbit trapped during
some limited interval of time, and non-trapped orbits.

Returning to Figures~\ref{fig1} and \ref{fig2}, of the red-circled points belonging to a
group, those with a blue or red cross represent stars with orbits
trapped by resonances during a time interval greater, or less than
5 Gyr, respectively. Group points without a cross represent orbits
not trapped by resonances. The division at 5 Gyr is arbitrary, just to
show long and short trapping. Some points to notice from these figures
are:

\begin{enumerate}
\item Resonant trapping may occur in or near the stable parts of a
resonant family.
\item There are group points which may be very close to a
\textit{stable} part of a resonant family but are \textit{not} trapped.
In a given Galactic potential, general conditions for a 3D orbit
to be trapped by a 2D resonance on the Galactic plane remain to be
analyzed. In Section \ref{orbnatr} some comments on non-trapped orbits
are given.
\item Resonant trapping of a given star in a group depends on the value
of $\Omega_b$. As noted above, the distribution of points in a
characteristic energy--J diagram varies strongly with the value of this
parameter.
\end{enumerate}

In Table~\ref{tabla3} we give a summary of the results shown in
Figures~\ref{fig1} and \ref{fig2}. For each star in a group, numbered as in
Table~\ref{tabla1}, and for each value of $\Omega_b$, the trapping
family and the trapping time in units of Gyr are given in each entry.
Empty spaces correspond to non-trapped orbits. Where ``other''
appears for the trapping family, it means that there are few missed
resonant families which were not computed in this study and in Paper I;
they are not very relevant in the results. The trapping families
III$\perp$ and VI$\perp$ denote the resonant families perpendicular
to families III and VI given in Paper I (its figures 12 and 15), i.e.,
those families rotated 90$^{\circ}$. Examples of perpendicular
families already appeared in Paper I: IV and V, VIII and IX, X and XI.
In regions where the points of groups G18-39 and G21-22 distribute,
families IV, VIII, and X have unstable parts but families V, IX, and
XI have stable parts, thus in these regions the later are the possible
trapping families. Related to this, the trapped points with a blue cross
in frame (b) of Figure~\ref{fig1} which lie close to \textit{unstable} parts of
curves III and VI (these points correspond to stars 19 and 17 in
group G18-39) are actually trapped by near \textit{stable} parts of
families III$\perp$ and VI$\perp$ (not shown in the figures).\\

\noindent Our main conclusions with the results in
Figures~\ref{fig1} and \ref{fig2} are:

\begin{enumerate}
\item The majority of stars in both groups G18-39 and G21-22 are
trapped by 2D resonances on the Galactic plane generated by the
Galactic bar.
\item There is not a single trapping family in each group; the main
trapping families in both groups are the families V and IX, with family
IX dominating in G21-22. Thus, it appears that the majority of stars
in both groups are members of the supergroups of stars in the Galaxy
trapped by the resonant families V and IX.
\item In the group G18-39 the interval 45--55
$\mathrm{km\,s^{-1}\,kpc^{-1}}$ in $\Omega_b$ produces the major
number of trapped orbits by families V and IX, and in the group G21-22
the corresponding interval is 45--60 $\mathrm{km\,s^{-1}\,kpc^{-1}}$.
\item For a given star in a group, a possible resonant trapping depends
strongly on the value of $\Omega_b$.
\end{enumerate}

\tabletypesize{\scriptsize}
\tablewidth{0pt}
\tablecolumns{6}
\begin{deluxetable}{cccccc}
\tablecaption{Trapping family and trapping time during
10 G\lowercase{yr} \label{tabla3}}
\tablehead{\\
\colhead{} & \multicolumn{5}{c}{$\Omega_b\, (\mathrm{km\,s^{-1}\,kpc^{-1}})$} \\
\cline{2-6}
 Number  & 40 & 45 & 50 & 55 & 60 \\
}
\startdata
\cutinhead{GROUP G18-39}
 1 & VI$\perp$;4 & ---          & IX;5.5        & IX;10       & ---   \\
 2 & ---         & ---          & V;10          & V;10        & V;10  \\
 3 & ---         & IX;10        & IX;8          & ---         & ---   \\
 4 & ---         & V;10         & V;10          & ---         & ---   \\
 5 & ---         & V;10         & V;10          & ---         & IX;2.5\\
 6 & ---         & $\sim$VI;4   & ---           & IX;8.5      & IX;4.5\\
 7 & V;10        & IX;4         & IX;10         & IX;10       & IX;10 \\
 8 & ---         & ---          & V;4.5         & V;10        & V;4   \\
 9 & V;10        & ---          & X;4           & IX;10       & IX;10 \\
 10 & V;10       & V;10         & VI$\perp$;3.5 & VI$\perp$;4 & IX;10 \\
 11 & ---        & V;10         & V;10          & V;10        & ---   \\
 12 & ---        & other;4      & IX;10         & IX;10       & ---   \\
 13 & V;10       & IX;3.5       & IX;5          & IX;10       & IX;8  \\
 14 & ---        & V;10         & V;10          & V;7         & ---   \\
 15 & ---        & V;6          & VI;3          & other;10    & ---   \\
 16 & VII;10     & IX;10        & IX;10         & ---         & ---   \\
 17 & V;10       & VI$\perp$;6  & other;4.5     & IX;5        & IX;10 \\
 18 & ---        & V;10         & V;10          & VI$\perp$;2 & IX;6  \\
 19 & ---        & III$\perp$;5 & V;10          & V;10        & V;5   \\
 20 & other;2.5  & III$\perp$;2 & ---           & V;10        & V;4   \\
 21 & VII;5      & IX;8         & IX;5          & ---         & XI;10 \\
 22 & other;10   & V;5          & V;4           & ---         & ---   \\
 23 & ---        & other;4      & V;10          & V;10        & ---   \\
 24 & V;4        & ---          & ---           & IX;4        & IX;8  \\
 25 & V;10       & V;8          & ---           & ---         & IX;4.5\\
\cutinhead{GROUP G21-22}
 1 & V;10        & ---          & ---           & IX;5        & IX;10 \\
 2 & V;10        & VI;3         & ---           & IX;10       & IX;8  \\
 3 & ---         & ---          & ---           & $\sim$XI;9.5 & ---  \\
 4 & ---         & V;10         & V;10          & ---         & ---   \\
 5 & ---         & ---          & ---           & ---         & ---   \\
 6 & IX;6        & IX;10        & IX;2.5        & ---         & XI;5  \\
 7 & V;10        & V;8      & $\sim$VI$\perp$;3 & ---         & IX;10 \\
 8 & VI;4        & IX;8         & IX;10         & IX;10       & IX;5  \\
 9 & VI;4        & IX;6         & IX;6          & IX;10       & IX;10 \\
 10 & ---        & IX;8         & IX;10         & IX;8        & IX;3.5\\
 11 & ---        & IX;8         & IX;10         & IX;10       & ---   \\
 12 & V;10       & ---          & IX;2.5        & IX;10       & IX;10 \\
\enddata
\end{deluxetable}

\subsection{Orbits}
\label{orbitas}

\begin{figure*}
\gridline{\fig{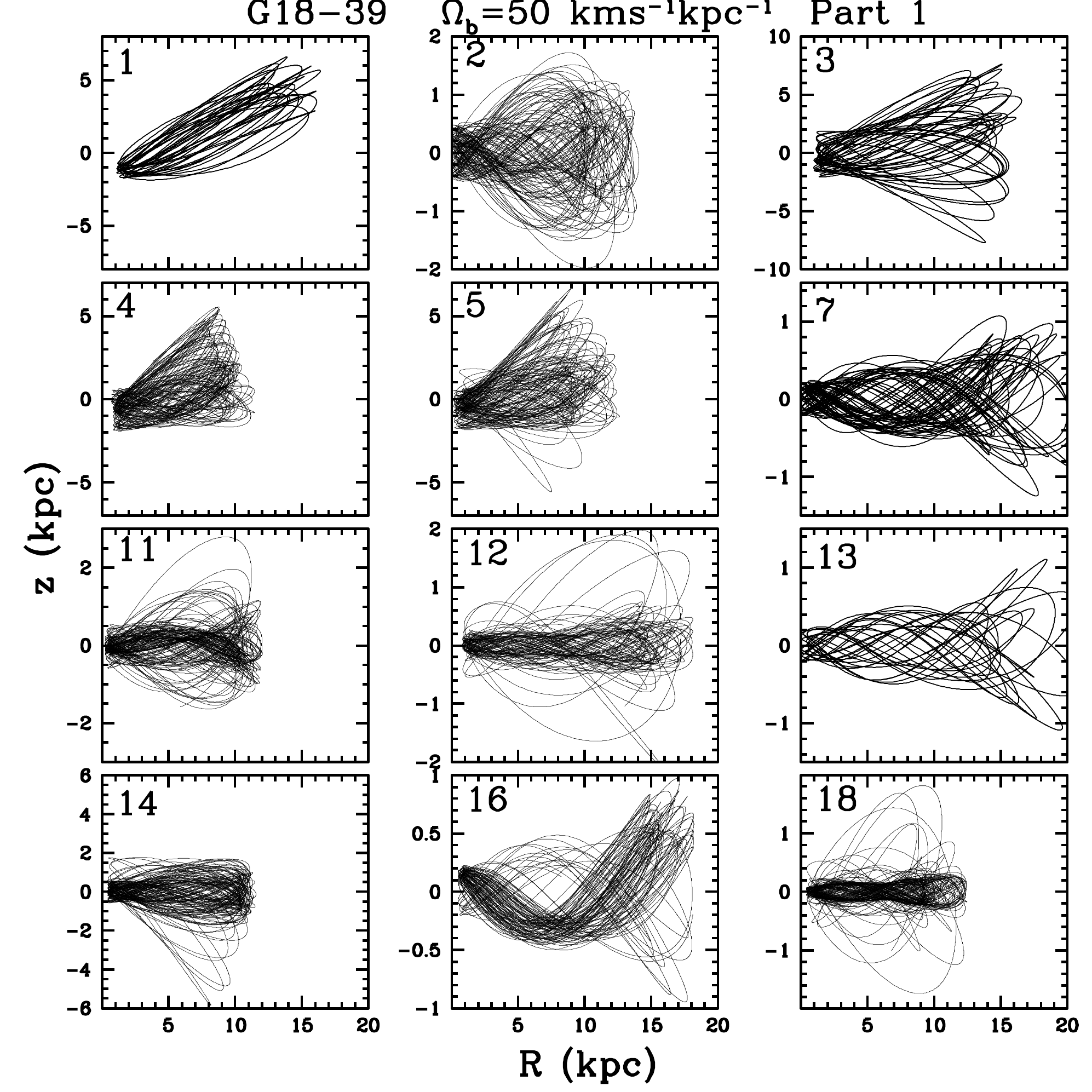}{0.5\textwidth}{(a)}
          \fig{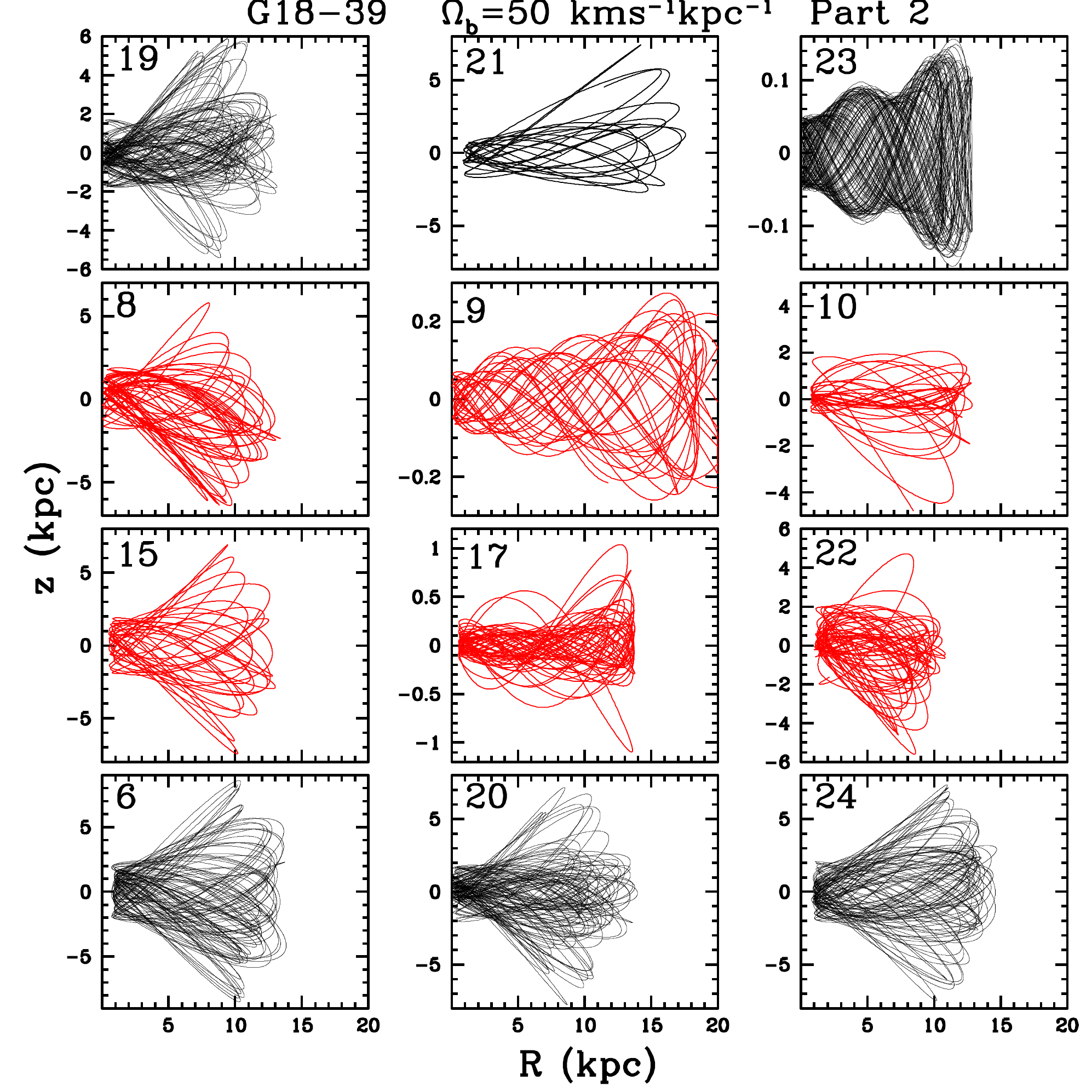}{0.5\textwidth}{(b)}}
\gridline{\fig{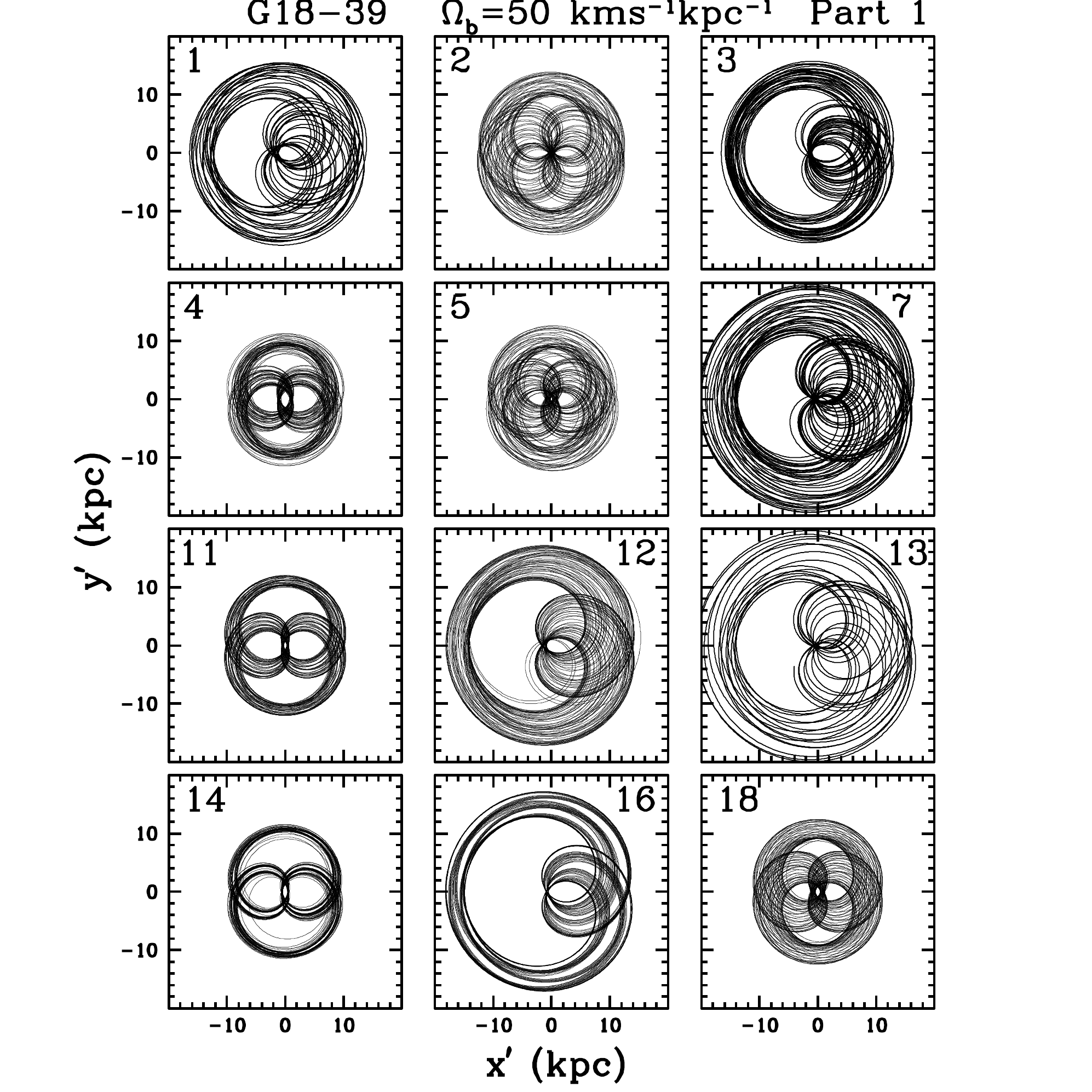}{0.5\textwidth}{(c)}
          \fig{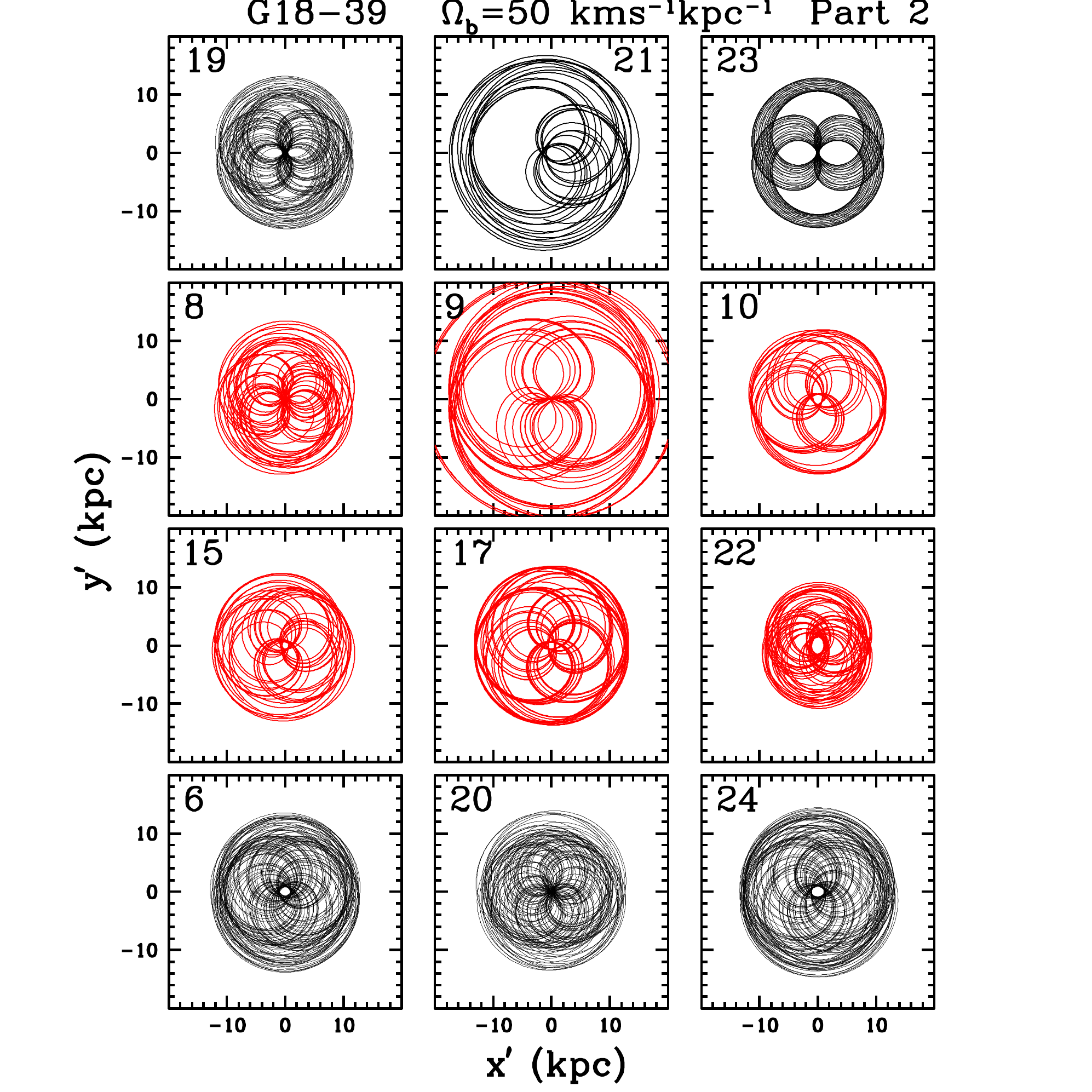}{0.5\textwidth}{(d)}}
\caption{Meridional (frames (a),(b)) and corresponding projected orbits
on the Galactic plane in the
non-inertial reference frame of the bar (frames (c),(d)), for 24 stars
in the group G18-39 with $\Omega_b = 50 ~\mathrm{km\,s^{-1}\,kpc^{-1}}$.
The star number is given inside each small frame. The orbits are shown
in the trapping intervals of time.
The twelve orbits in frames (a),(c) and the three orbits at the top
in frames (b),(d) are trapped in an interval of time
greater or equal to 5 Gyr. 
The six orbits in red color in frames (b),(d) are trapped in an
interval of time less
than 5 Gyr, and the last three in these frames are non-trapped orbits.
\label{fig7}}
\end{figure*}

\begin{figure*}
\gridline{\fig{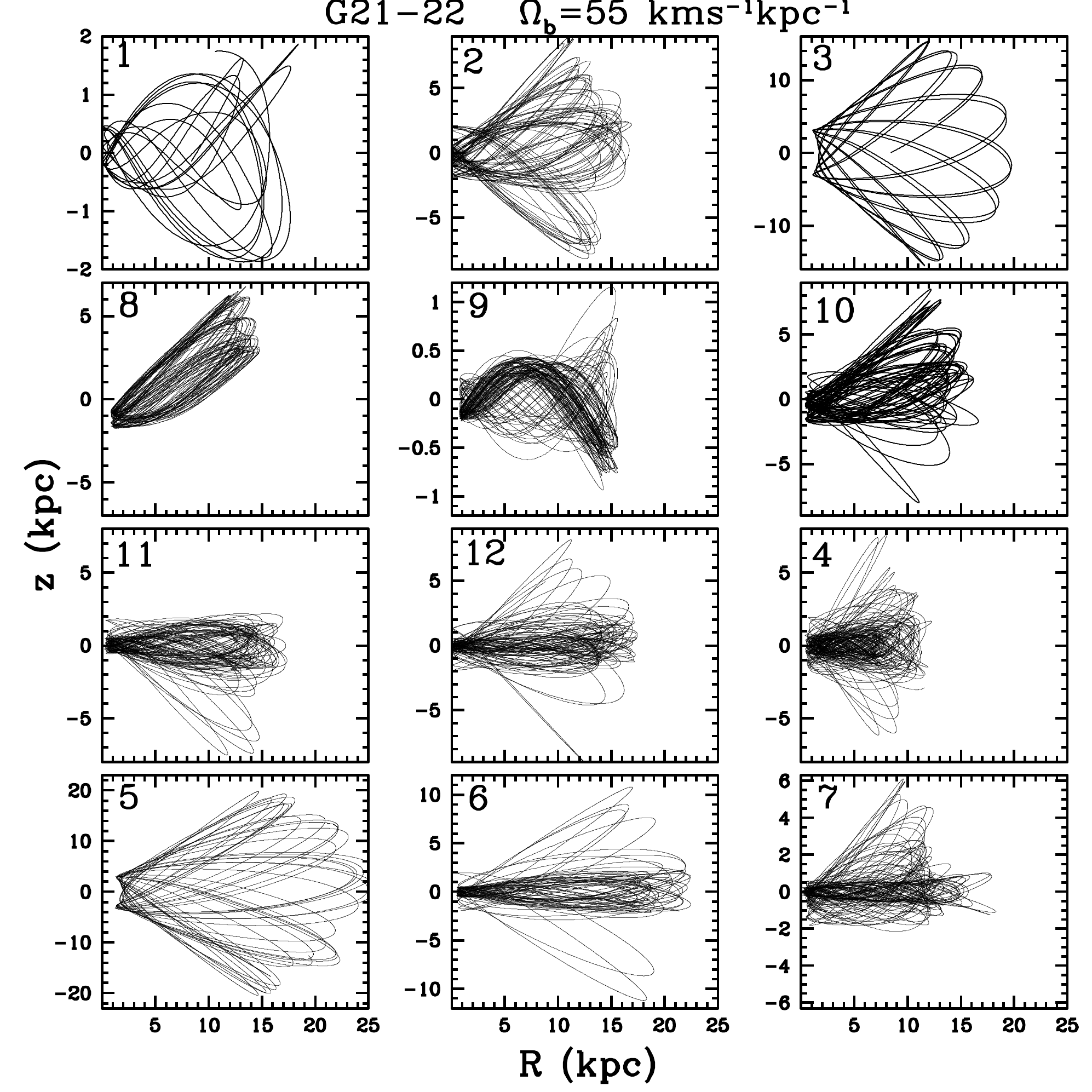}{0.5\textwidth}{(a)}
          \fig{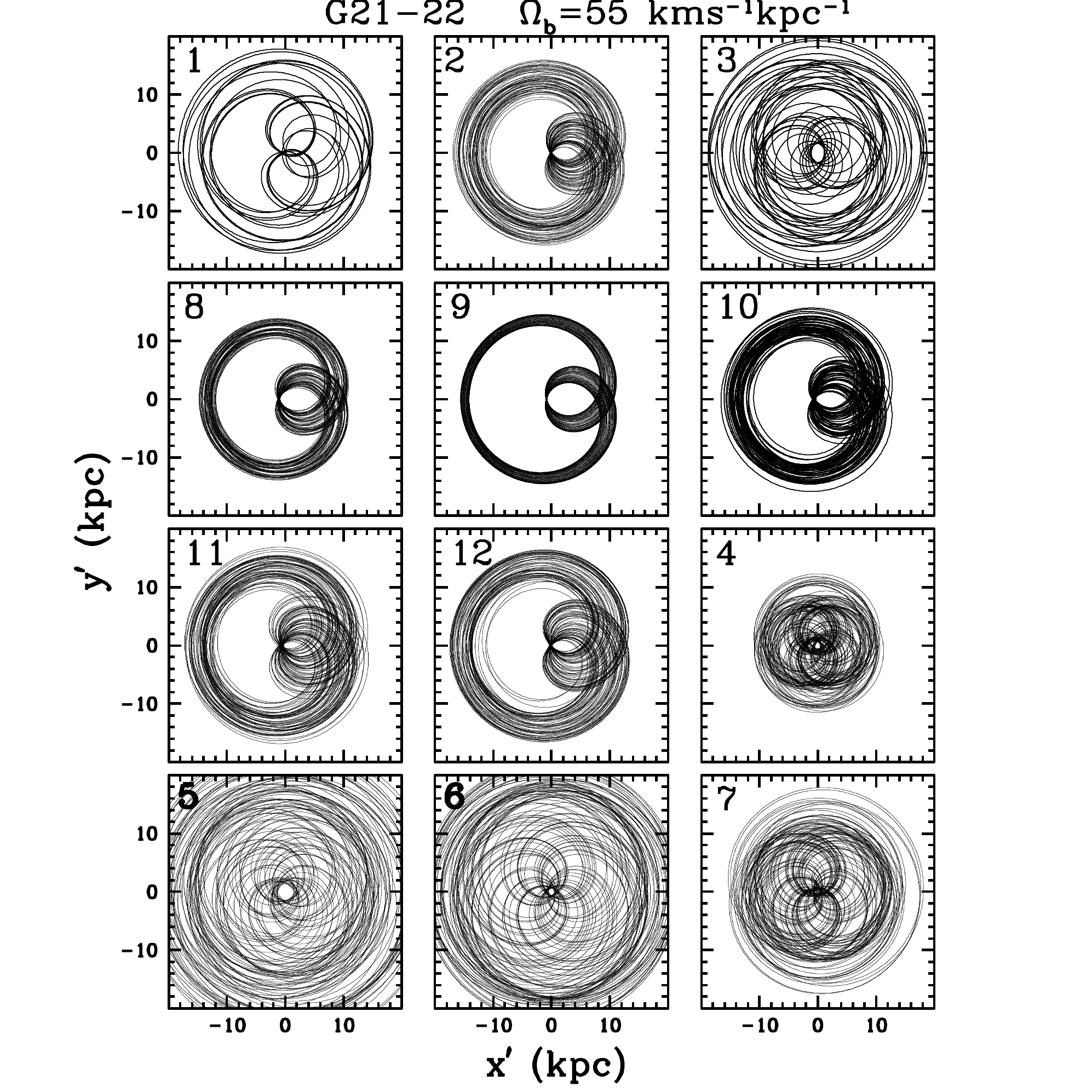}{0.5\textwidth}{(b)}}
\caption{As in Figure~\ref{fig7}, for the twelve stars in the group
G21-22 with $\Omega_b = 55 ~\mathrm{km\,s^{-1}\,kpc^{-1}}$.
The meridional and projected orbits are given in frames (a) and (b),
respectively. The first eight stars have trapped orbits in an 
interval of time greater or equal to 5 Gyr. The last four stars 
have non-trapped orbits. \label{fig8}} 
\end{figure*}

In this part we present some examples of star orbits in the two groups
G18-39 and G21-22. Figure~\ref{fig7} shows the orbits of 24 of the
25 stars in the group G18-39, with
$\Omega_b = 50 ~\mathrm{km\,s^{-1}\,kpc^{-1}}$. The meridional orbits
are shown in frames (a) and (b), and the corresponding projected orbits
on the Galactic plane in the non-inertial reference frame of the bar,
are shown in frames (c) and (d). The star number is given inside each
small frame. The twelve stars in frames (a),(c) and the
first three in frames (b),(d) have orbits trapped in an interval of
time greater or equal to 5 Gyr. See the corresponding column in
Table~\ref{tabla3} for the trapping family and trapping time. Notice
the different maximum z-distances reached in each orbit. The six orbits
in red color in frames (b),(d) are approximately trapped in an
interval of time less than 5 Gyr, and the last three in these frames 
are non-trapped orbits. 
For the group G21-22, Figure~\ref{fig8} gives the
meridional (frame (a)) and projected (frame (b)) orbits of the 12 stars
in this group, with
$\Omega_b = 55 ~\mathrm{km\,s^{-1}\,kpc^{-1}}$. The first eight stars
have trapped orbits in an interval of time greater or equal to 5 Gyr.
The last four stars have non-trapped orbits. In this group and with the
given value of $\Omega_b$, the main trapping family is family IX.

\subsection{Monte Carlo simulations}
\label{mc}

\begin{figure*}
\gridline{\fig{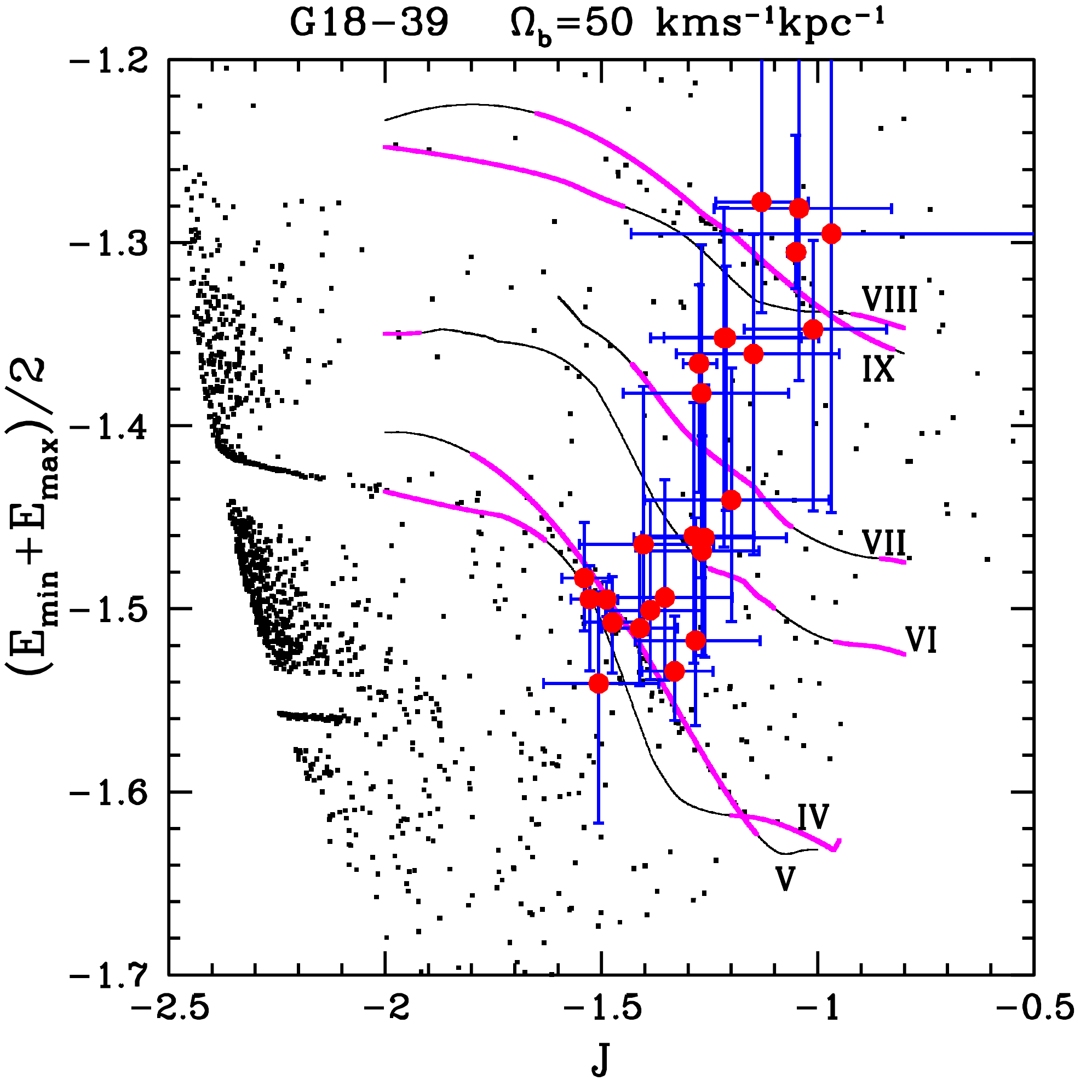}{0.5\textwidth}{(a)}
          \fig{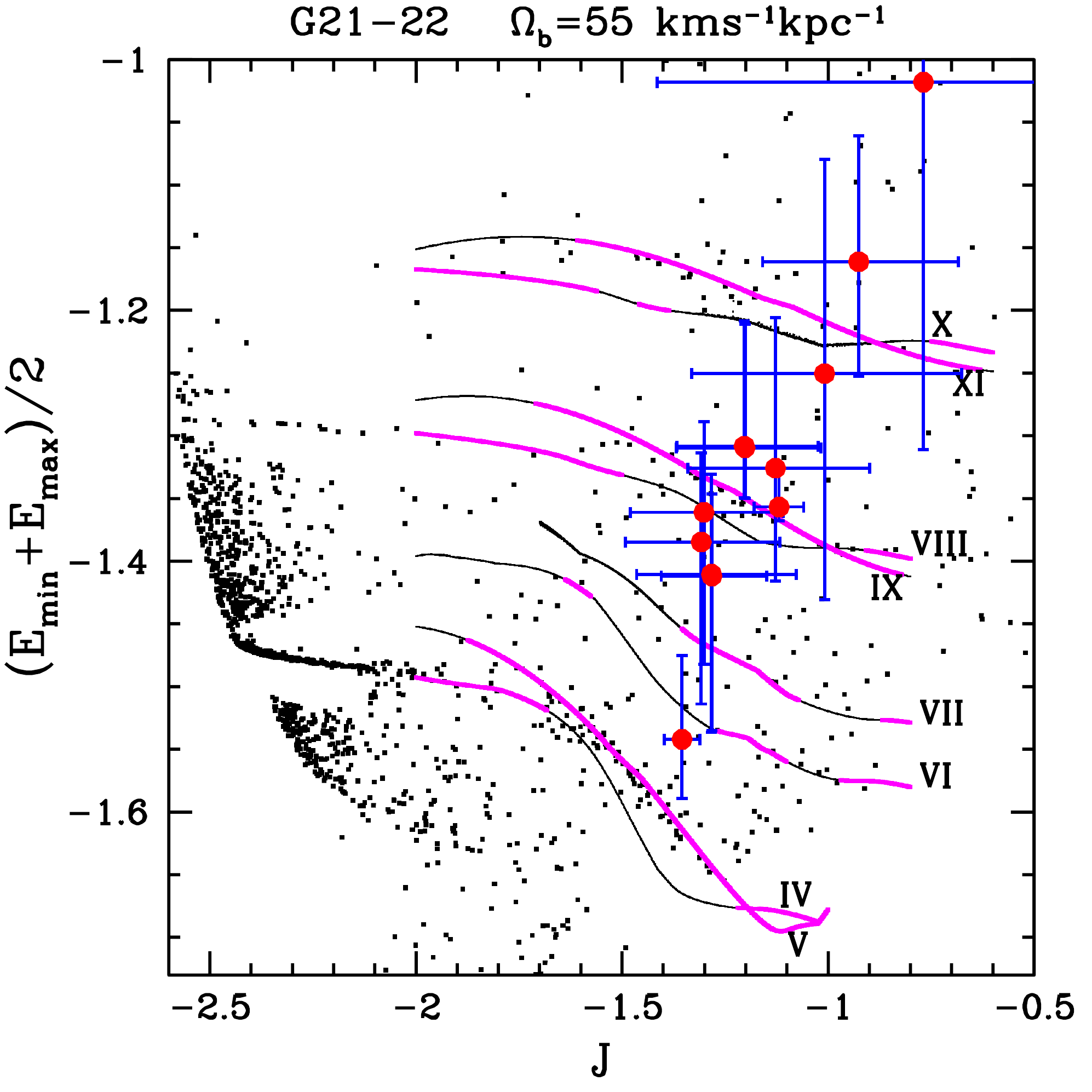}{0.5\textwidth}{(b)}}
\caption{Monte Carlo simulations, (a) in the group G18-39 with
$\Omega_b = 50~\mathrm{km\,s^{-1}\,kpc^{-1}}$, and (b) in the group
G21-22 with $\Omega_b = 55 ~\mathrm{km\,s^{-1}\,kpc^{-1}}$.
The red points are the resulting average positions and the blue
lines their associated dispersions. Compare
with frames (c) and (d) of Figures~\ref{fig1} and \ref{fig2},
respectively. \label{fig9}}
\end{figure*}

To see the effect of the uncertainties in the kinematic parameters of
the stars in both groups G18-39 and G21-22, listed in
Table~\ref{tabla1}, two representative Monte Carlo simulations are
given in this subsection. For each star, 500 orbits were computed with
initial conditions drawn from the set of varying kinematic parameters,
including the Solar motion, with $R_0$, $\Theta_0$ of the LSR and
$\Omega_b$ of the bar kept fixed. Figure~\ref{fig9} 
shows in frame (a) the results with
$\Omega_b = 50~\mathrm{km\,s^{-1}\,kpc^{-1}}$
in the group G18-39, and in frame (b) the results with
$\Omega_b = 55~\mathrm{km\,s^{-1}\,kpc^{-1}}$
in the group G21-22. The red points are the resulting average positions
and the blue lines their associated dispersions. Comparing with the
results obtained in frame (c) of Figure~\ref{fig1} and frame (d) of
Figure~\ref{fig2} using the mean
values of the parameters, approximate accumulations around resonant
families V and IX are recovered, as was the conclusion in previous
subsections.

\section{The two moving groups in the Bottlinger diagram}
\label{pUV}

\subsection{The trapped orbits}
\label{orbatr}

In the previous section we concluded that the majority of stars in
the two groups G18-39 and G21-22 are trapped by the two main resonant
families V and IX on the Galactic plane, generated by the Galactic bar.
This must be reflected in their velocity field parallel to the Galactic
plane observed in the solar neighborhood. In this section the two
groups are analyzed in the U$^{\prime}$--V$^{\prime}$ plane, or
Bottlinger diagram, with
U$^{\prime}$,V$^{\prime}$ velocities with respect to the LSR.
We compare their observed distribution in this plane with the one that
can be inferred from the kinematic properties of their trapping
families.

\begin{figure}[t!]
\plotone{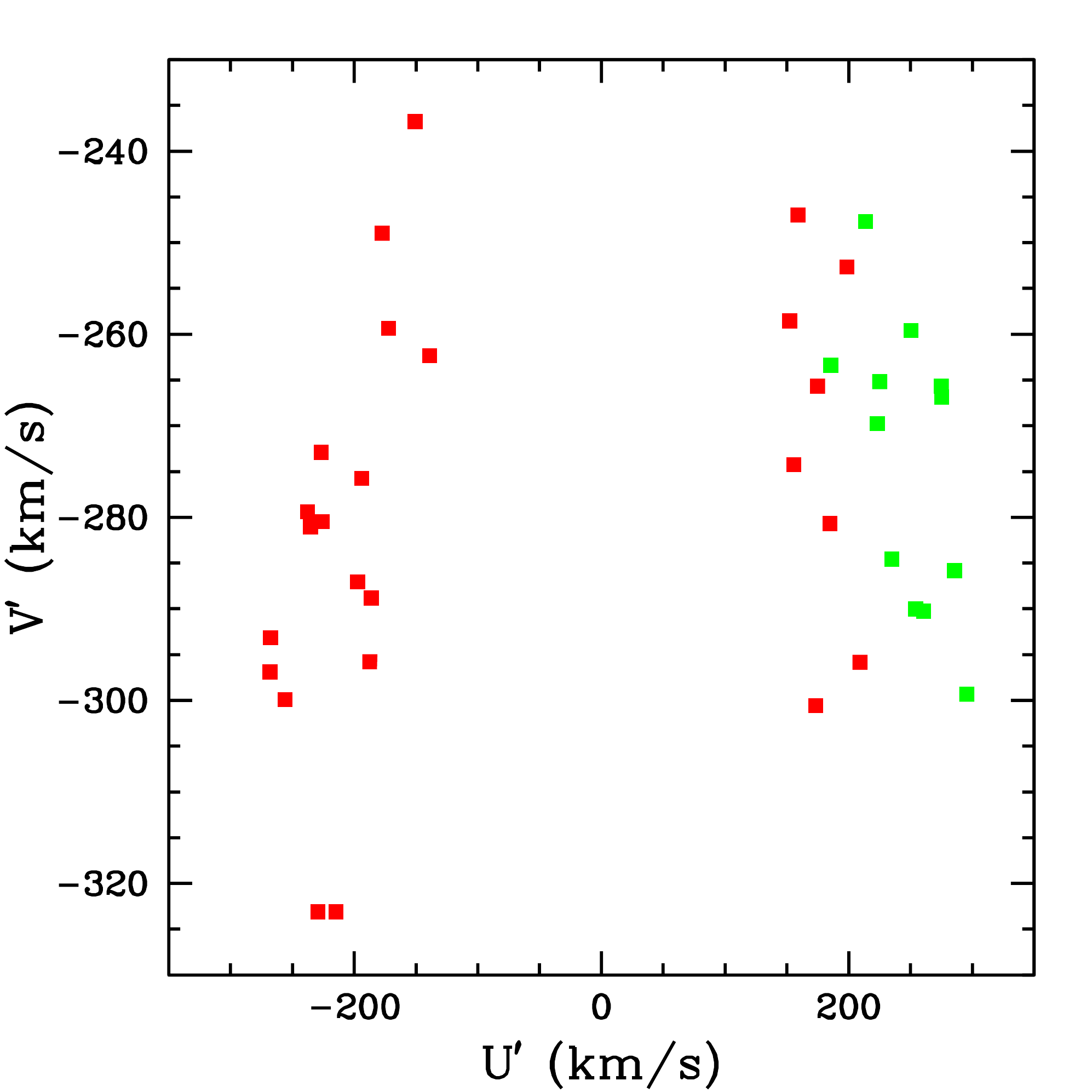}
\caption{The two groups in the Bottlinger diagram,
with U$^{\prime}$,V$^{\prime}$ observed velocities with respect to
the LSR. The red points correspond to G18-39 and the green points to
G21-22. \label{fig10}}
\end{figure}

In Figure~\ref{fig10} we give the observed distribution of the two
groups in the Bottlinger diagram, obtained with the
updated kinematic parameters listed in Table~\ref{tabla1}. The
colors used to differentiate the groups are the same as in figure 7
in the original analysis of the groups made by
\citet{2012RMxAA..48..109S}. Red color for the group
G18-39, with a double structure in the U$^{\prime}$ velocity, and green
color for G21-22, only appearing at U$^{\prime}$ positive, shifted
slightly to the right of the G18-39 structure on this side.

To model the distributions shown in Figure~\ref{fig10} we analize the
velocity field U$^{\prime}$,V$^{\prime}$ in the solar neighborhood
generated by the two main trapping families V and IX. The following
procedure is considered: to illustrate the method, we focus on the
results shown in frames (c) and (d) of Figures~\ref{fig1} and
\ref{fig2}, respectively; i.e., the cases
$\Omega_b = 50 ~\mathrm{km\,s^{-1}\,kpc^{-1}}$ in the group G18-39,
and $\Omega_b = 55 ~\mathrm{km\,s^{-1}\,kpc^{-1}}$ in the group G21-22.
In these two cases, the star orbits in both groups trapped by families
V and IX lie in some known different intervals in $J$, the Jacobi
constant. Taking a fine grid in $J$, 2D periodic orbits of a trapping
family lying in each of these intervals are computed. Also, some
different 2D tube orbits around these periodic orbits are computed
during several revolutions around the Galactic center. Tube orbits
have already been employed in the extended Poincar\'e diagrams shown
in Figures~\ref{fig4} and \ref{fig5}. In those figures we saw that a
3D trapped orbit approximately distributes among the tube orbits; thus,
we expect that the 2D velocity field produced by tube orbits reflects
on the observed distribution of U$^{\prime}$,V$^{\prime}$ velocities
of 3D trapped star orbits. To proceed with the analysis, a radius of
500 pc is considered for the solar vicinity, which is an appropriate
value according to the distances given in Table~\ref{tabla1} for the
stars in both groups. Next, with a given
\textit{present} position angle of the major axis of the Galactic bar
and the Sun's galactocentric distance $R_0$ (see Table~\ref{tabla2}),
the solar neighborhood is defined in the non-inertial reference frame
where the bar is at rest, in which the 2D periodic and tube orbits are
computed. When one of these orbits crosses the solar neighborhood,
we take orbital points within this vicinity separated by 100 pc, and in
 each point
transform the velocity to the inertial reference frame; finally, the
corresponding U$^{\prime}$,V$^{\prime}$ velocities with respect to the
LSR are obtained in all these internal points, and a comparison with
the observed distributions can be made.

\begin{figure*}
\gridline{\fig{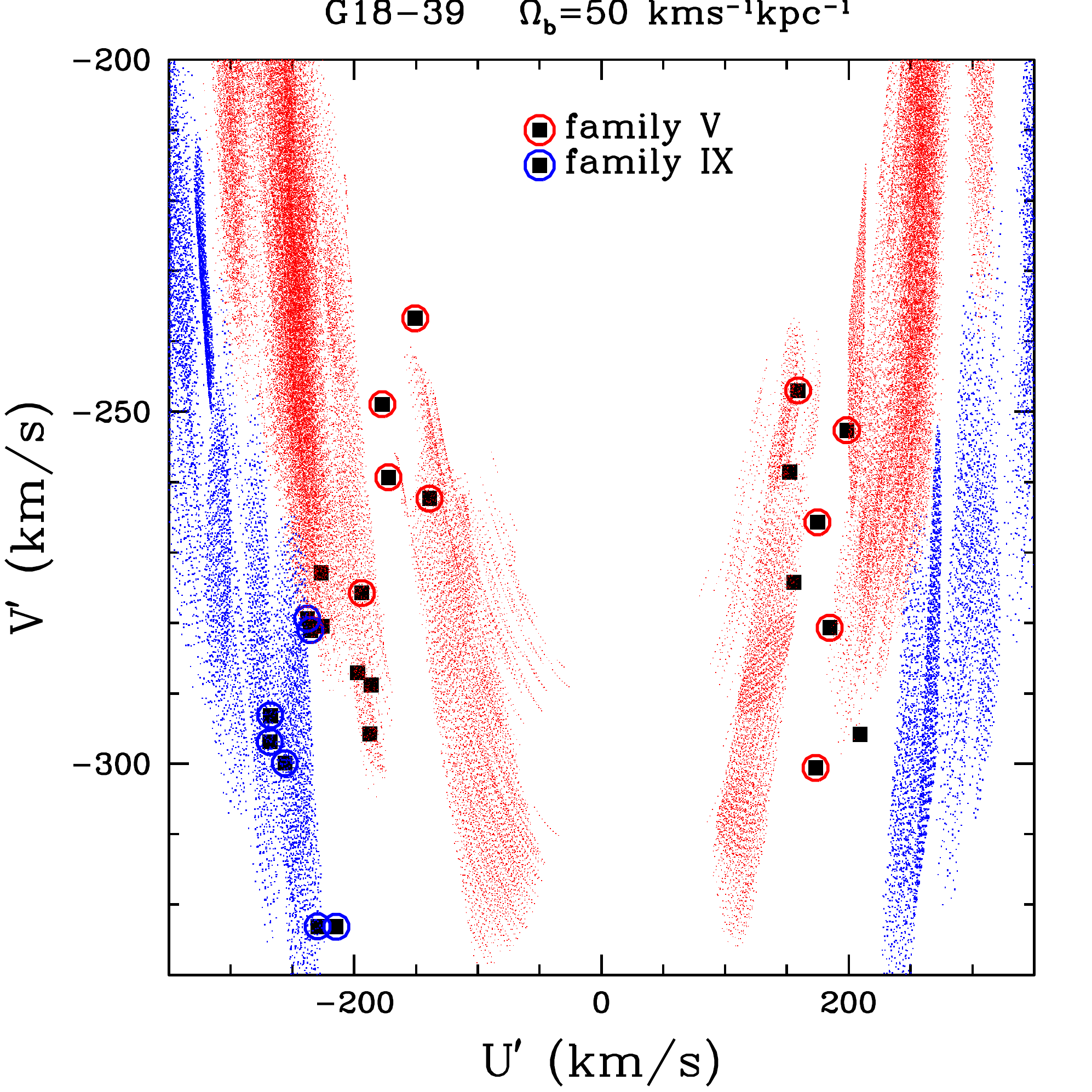}{0.5\textwidth}{(a)}
          \fig{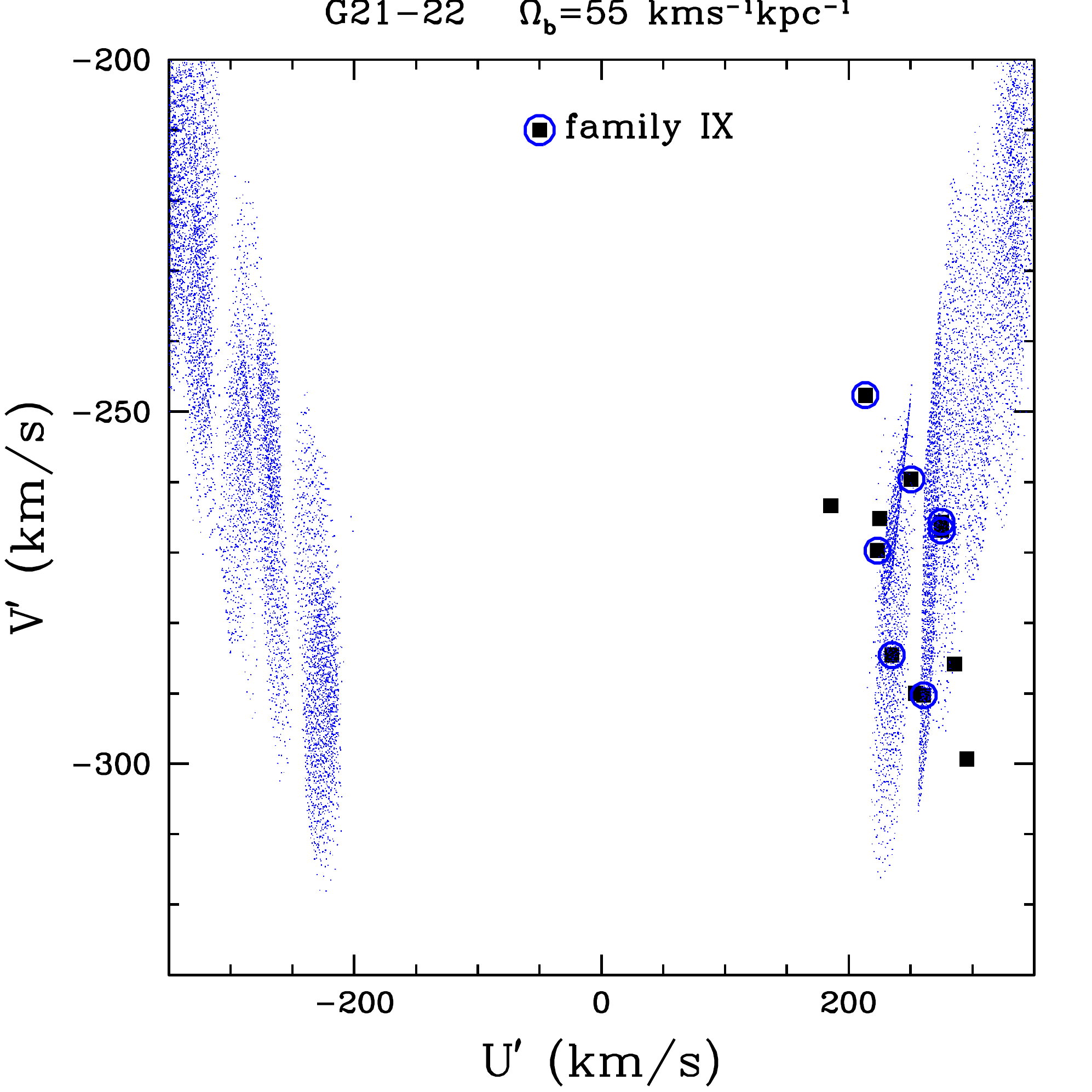}{0.5\textwidth}{(b)}}
\caption{The observed and inferred distributions in the
Bottlinger diagram for the group G18-39 with
$\Omega_b = 50 ~\mathrm{km\,s^{-1}\,kpc^{-1}}$ (frame (a)), and
the group G21-22 with 
$\Omega_b = 55 ~\mathrm{km\,s^{-1}\,kpc^{-1}}$ (frame (b)).
In (a) the black points
correspond to the observed distribution in Figure~\ref{fig10} for the
group G18-39. The points with a red circle are the stars trapped by
 family V,
and those with a blue circle are stars trapped by family IX. The red
and blue regions show values for internal orbital points in the solar
neighborhood of periodic and tube orbits in families V and IX,
respectively. There is an approximate concordance of the circled
red and blue points with the corresponding red and blue regions.
In (b) the black points are the observed green points in
Figure~\ref{fig10} for the group G21-22. The points with a blue circle
are orbits trapped 
by family IX. The blue regions correspond to periodic and tube orbits
in family IX. \label{fig11}}
\end{figure*}

Figure~\ref{fig11} shows the comparisons for the two groups. 
In frame (a) the black points 
correspond to the observed distribution in Figure~\ref{fig10} for the
group G18-39; the points with a red circle are the stars trapped by
 family V,
and with a blue circle the stars trapped by family IX. The very small
red and blue points, which appear approximately as a continuum,
correspond respectively to internal orbital points in the solar
neighborhood of periodic and tube orbits in families V and IX. There
are some gaps in the distribution of these small points because the
number of tube orbits around each periodic orbit was small (about 10
orbits). Notice the approximate concordance of the circled red
and blue points with the corresponding red and blue regions. There is
a blue region at positive U$^{\prime}$ which is not populated by stars
in the group. This region appears again in frame (b), in
relation with the group G21-22. In this second frame the black points
are the green points in Figure~\ref{fig10}. From Table~\ref{tabla3}, the
main trapping family is family IX, with family V not appearing in this
case. Again, the trapped points, with a blue circle, lie approximately
in a blue region of periodic and tube orbits in family IX. The blue
regions here and in frame (a) have almost the same location,
the small difference is due to the different values
$\Omega_b = 50,55 ~\mathrm{km\,s^{-1}\,kpc^{-1}}$, respectively.
This Figure~\ref{fig11} already points to the
observed shift or separation of the groups shown in Figure~\ref{fig10}:
some stars in the group G18-39 are trapped by family V and lie on its
two branches at U$^{\prime}$ positive and negative; other stars in
this group are trapped by family IX and lie on the U$^{\prime}$-
negative branch of this family. The stars in the group G21-22 are
mainly trapped by family IX, lying on its U$^{\prime}$-positive branch.

\begin{figure}[t!]
\plotone{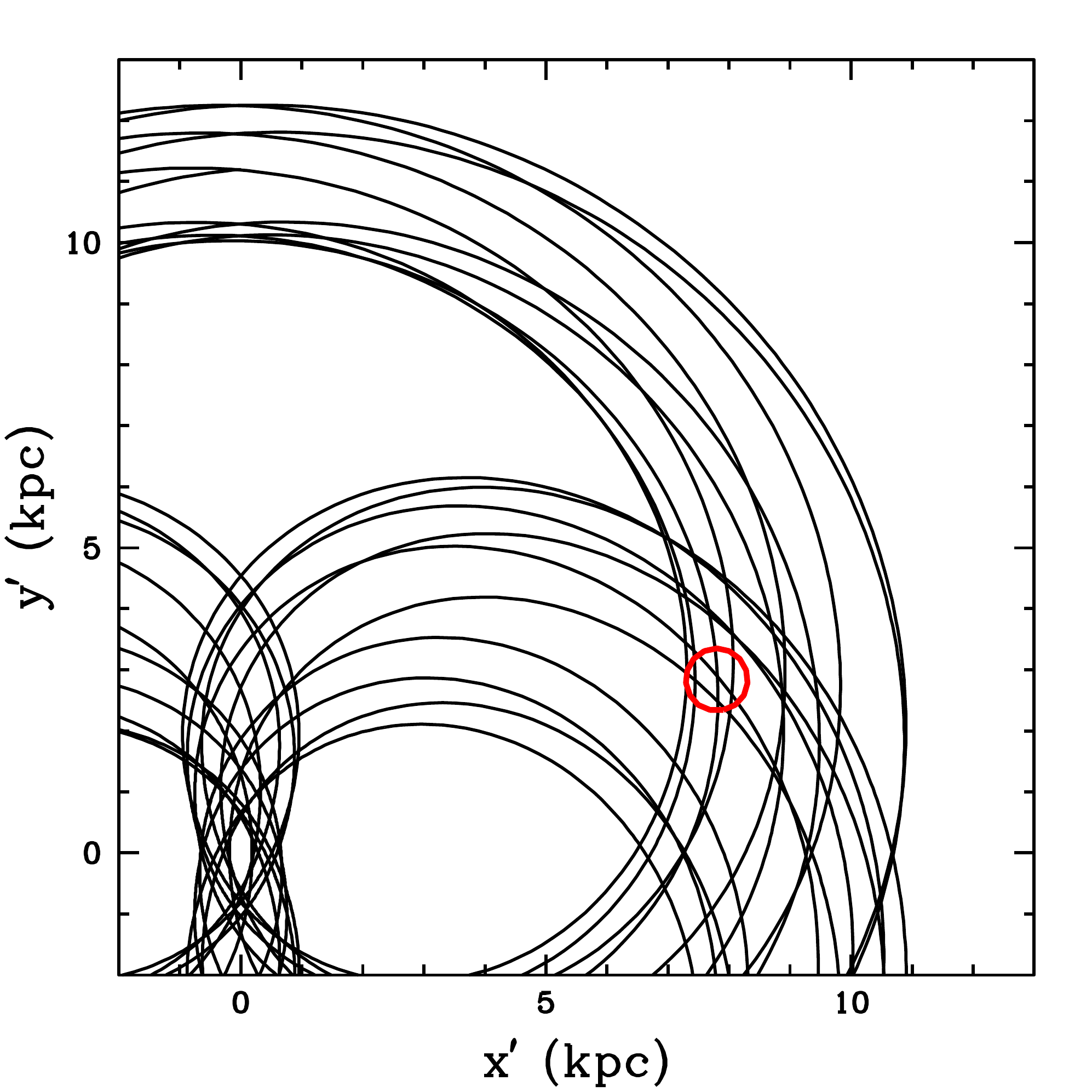}
\caption{This figure illustrates the appearance of two
branches in the Bottlinger diagram, at U$^{\prime}$
positive and U$^{\prime}$ negative as shown in Figure~\ref{fig11}, 
of periodic and tube orbits in families V and IX.
As an example, the tube orbit in family V given in frame (d) of
Figure~\ref{fig3} is plotted. The solar vicinity is shown with a red
circle at 20$^{\circ}$ with the $x^{\prime}$-axis, the major axis of
the bar. When the orbit crosses the solar vicinity the U$^{\prime}$
velocity has positive and negative values within narrow intervals, due
to ordered motion of the tube orbit. A similar figure can be done for
family IX. \label{fig12}}
\end{figure}

The appearance of two branches, shown in Figure~\ref{fig11}, 
one at U$^{\prime}$ positive and the other at U$^{\prime}$
negative, due to periodic and tube orbits in families V and IX,
results from the orbital behavior of these families when they cross
the solar neighborhood. To illustrate this point, in Figure~\ref{fig12}
we show again the tube orbit in family V given in frame (d) of
Figure~\ref{fig3}. The solar vicinity is shown with a red circle and
the major axis of the bar, i.e., the $x^{\prime}$-axis, makes an
angle of 20$^{\circ}$ with the Sun-Galactic center line. Due to ordered
motion of tube orbits, in the U$^{\prime}$ velocity the orbit has
positive and negative values in somewhat narrow intervals when
it crosses the solar vicinity. Different tube orbits will have
different intervals, thus their superposition will enlarge the total
range of values. A similar figure can be done for family IX, taking for
example a tube orbit with the form of the projected orbit of star
number 10 in frame (b) of Figure~\ref{fig8}. The position of the two
 branches in
each family in the Bottlinger diagram will be different.

\begin{figure*}
\gridline{\fig{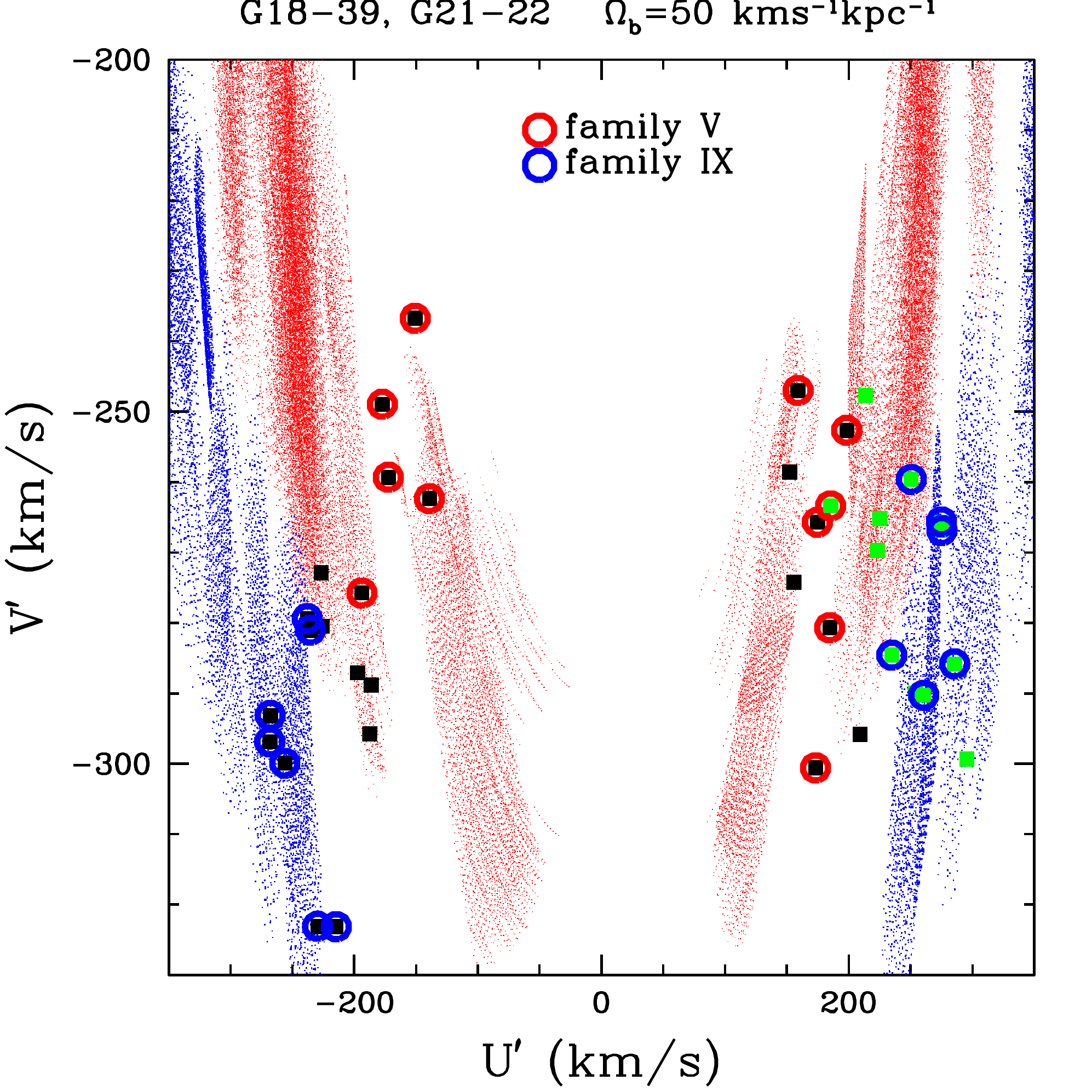}{0.5\textwidth}{(a)}
          \fig{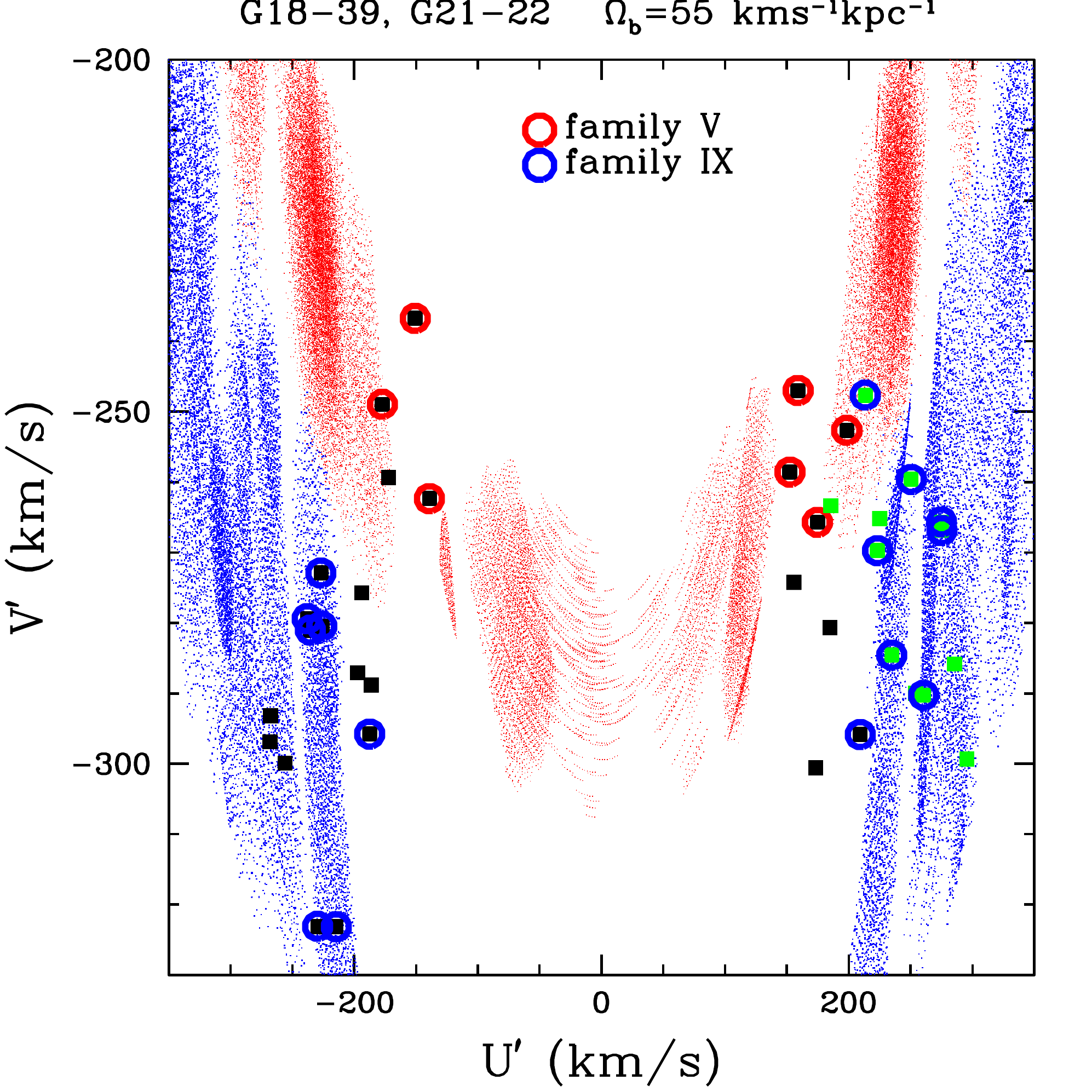}{0.5\textwidth}{(b)}}
\caption{The two groups G18-39 and G21-22 in the Bottlinger diagram
with $\Omega_b = 50~\mathrm{km\,s^{-1}\,kpc^{-1}}$ (frame (a)), and
$\Omega_b = 55~\mathrm{km\,s^{-1}\,kpc^{-1}}$ (frame (b)).
The black points correspond to the group G18-39 and the green points
to the group G21-22. Points circled in red are stars trapped by family
V, and
those circled in blue are stars trapped by family IX. All these circled
points lie approximately in the corresponding red and blue regions
produced by families V and IX, respectively. In frame (b) the number of
stars trapped by family V and family IX is respectively less and
greater than in frame (a). \label{fig13}}
\end{figure*}

To compare with the joint observed distribution in Figure~\ref{fig10},
in Figure~\ref{fig13} we show together the two groups
G18-39 and G21-22 under the two different values
$\Omega_b = 50, 55 ~\mathrm{km\,s^{-1}\,kpc^{-1}}$. In both frames
(a) and (b) the
black points correspond to the group G18-39 and the green points to the
group G21-22. Points circled in red are stars trapped by family V, and
those circled in blue are stars trapped by family IX. We note that
with $\Omega_b = 50 ~\mathrm{km\,s^{-1}\,kpc^{-1}}$ there are more
stars trapped by family V than with the higher value
$\Omega_b = 55 ~\mathrm{km\,s^{-1}\,kpc^{-1}}$, and conversely, the
number of stars trapped by family IX increases with the higher value
of $\Omega_b$ (see Table~\ref{tabla3}). Also, the red-circled and blue-circled points lie
approximately in the corresponding red and blue regions produced by
families V and IX, respectively; thus showing the observed shift in
Figure~\ref{fig10}.

\subsection{The non-trapped orbits}
\label{orbnatr}

In the last section the U$^{\prime}$,V$^{\prime}$ velocities in the
solar neighborhood of trapped 3D orbits in the two groups G18-39 and
G21-22 have been modeled through the 2D velocity fields of periodic and
tube orbits in the resonant families V and IX on the Galactic plane.
Of the uncircled points in Figures~\ref{fig11} and \ref{fig13},
several correspond to non-trapped orbits,
but lie on or near the regions of families V and IX; the rest
correspond to orbits trapped by other families.

\begin{figure}[t!]
\plotone{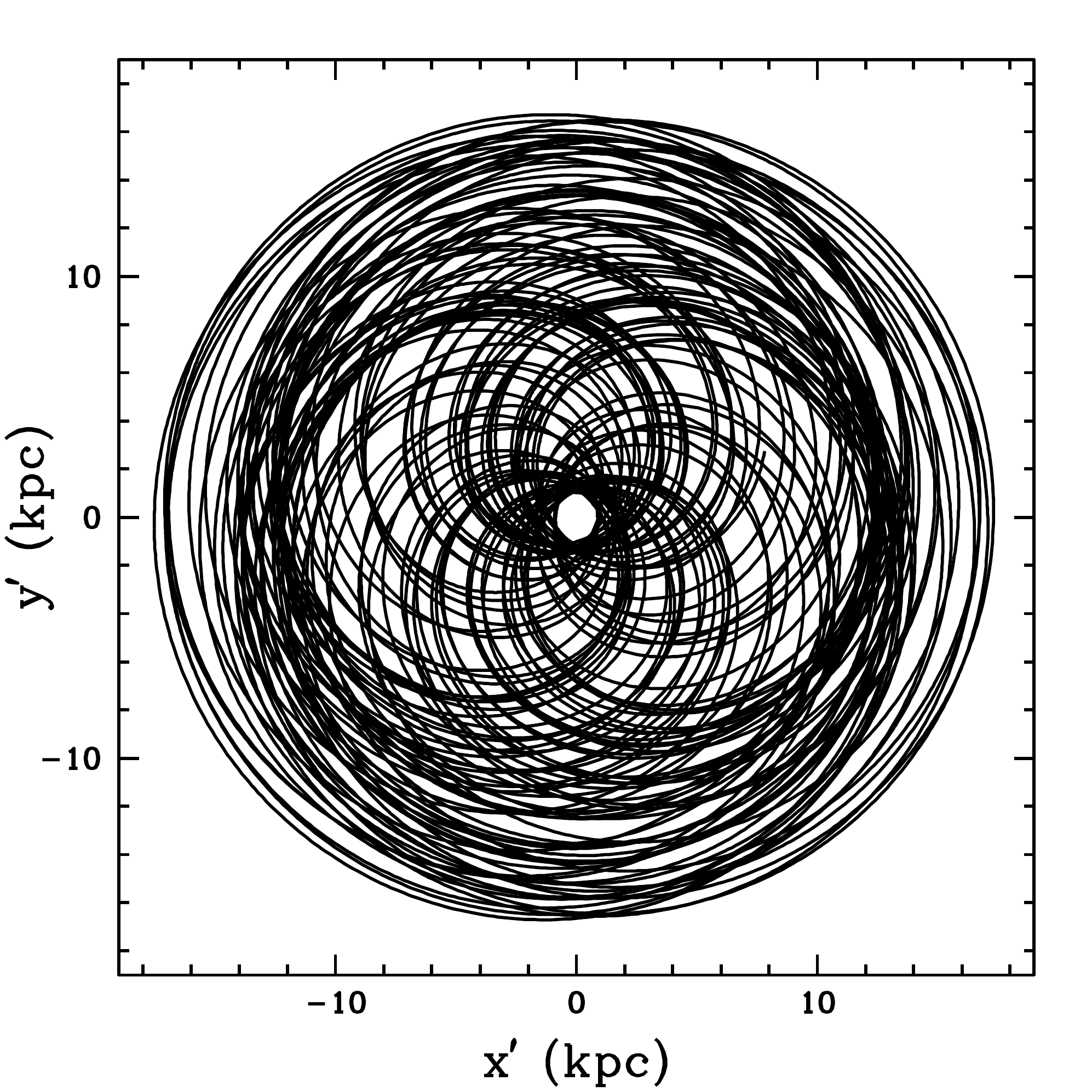}
\caption{This is the projected orbit during the total time of 10 Gyr
of star with number 21 in the group G18-39, with
$\Omega_b = 50~\mathrm{km\,s^{-1}\,kpc^{-1}}$. In an interval of time
of 5 Gyr, listed in Table~\ref{tabla3}, the orbit is trapped by family
IX. See first row in frame (d) of Figure~\ref{fig7} for the projected
orbit in this interval. \label{fig14}}
\end{figure}

For trapped orbits, in Section \ref{EJpl} we mentioned how the trapping
time was determined in a given orbit. In Table~\ref{tabla3} we have
listed these intervals of time in each case. Most of the trapped orbits
have trapping intervals of time equal to the total computed orbital
time of 10 Gyr. In Figure~\ref{fig14} we show an example of an orbit
which is trapped in a limited interval of time, but at first sight this
orbit appears as a non-trapped orbit. This Figure gives the projected
orbit, during the total time of 10 Gyr, of star with number 21 in the
group G18-39, with $\Omega_b = 50~\mathrm{km\,s^{-1}\,kpc^{-1}}$.
There is an interval of time (5 Gyr), listed in Table~\ref{tabla3},
 where the
orbit is trapped by family IX; see this part of the orbit in the
first row in frame (d) of Figure~\ref{fig7}. In this interval of time,
which includes the present time, the orbit keeps its form fixed in the
non-inertial frame, and outside this interval this form attains a
precession-like motion around the $z^{\prime}$ axis, like in a rosette
orbit, almost with no significant variation of its radial and azimuthal
motions in the inner Galactic region; this is only approximate, due to
the non-axisymmetry of the potential. This means that the orbital
U$^{\prime}$,V$^{\prime}$ velocity field within the solar vicinity will
remain approximately the same at later times, when the projected orbit
crosses this vicinity again, as that presented at the initial trapping
time interval. For non-trapped orbits close to a trapping region the
precession-like motion will be the dominant orbital behavior, and it
is expected that their points in the Bottlinger diagram when the
orbit is inside the solar vicinity will lie close to the trapped
orbits.

\begin{figure}[t!]
\plotone{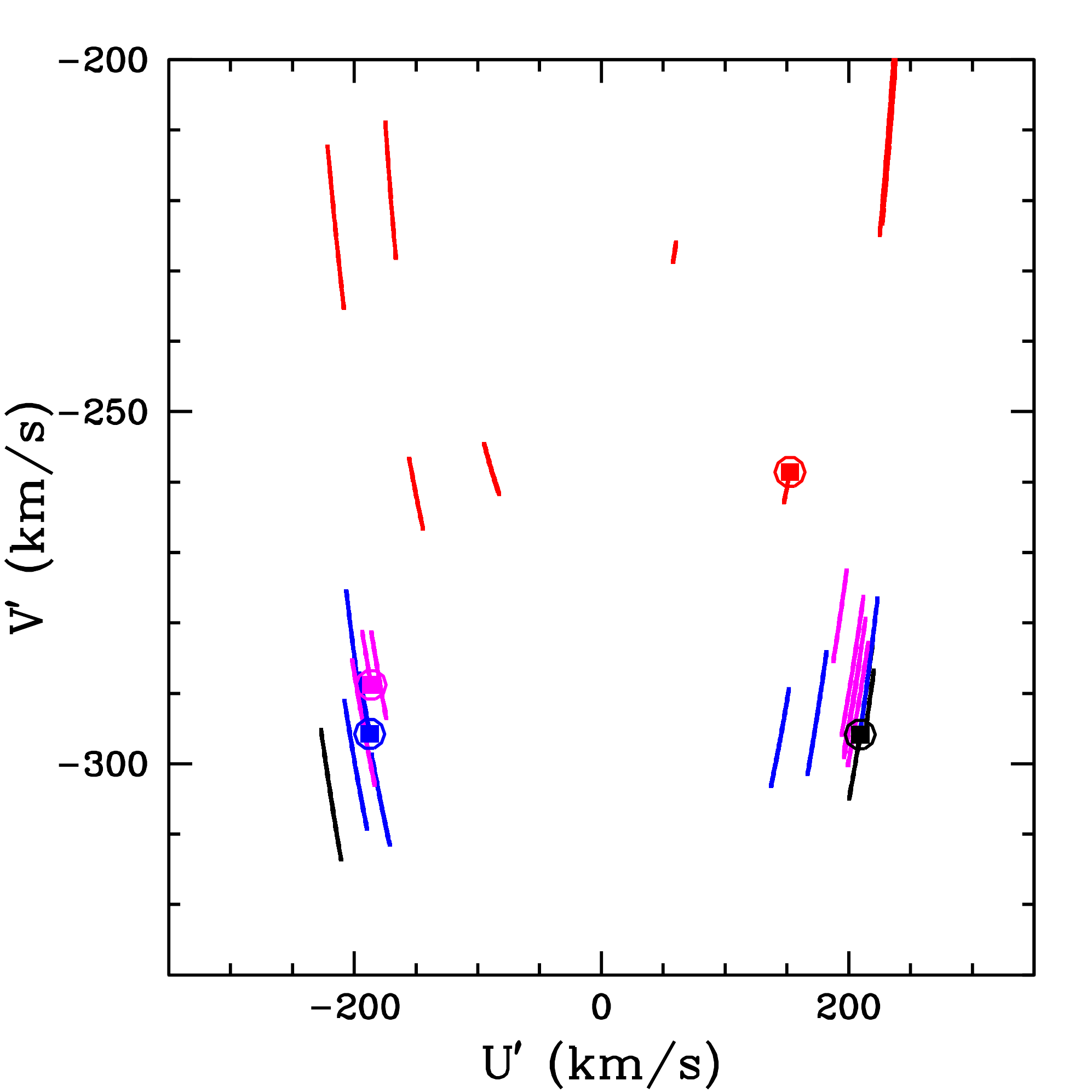}
\caption{The U$^{\prime}$,V$^{\prime}$ velocities for the four
non-trapped orbits of stars with numbers 6,20,24,25 in the group
G18-39 with $\Omega_b = 50~\mathrm{km\,s^{-1}\,kpc^{-1}}$. See
Table~\ref{tabla3}. The colored big circled squares for these stars are
four of the uncircled points in frame (a) of Figure~\ref{fig11}.
 The lines with
corresponding color define other velocities in projected positions
within the solar vicinity. These lines lie approximately in the red and
blue regions produced by families V and IX in frame (a) of
 Figure~\ref{fig11}, and
these four non-trapped orbits could be nearly trapped by these
families. \label{fig15}}
\end{figure}

To analyze this last point, for non-trapped orbits we calculate the 3D
orbit during 10 Gyr and determine the U$^{\prime}$,V$^{\prime}$
velocities in orbital points whose \textit{projected} positions on the
non-inertial $x^{\prime},y^{\prime}$ Galactic plane lie within the
solar vicinity. As an example, we consider the four non-trapped orbits
in the group G18-39 with $\Omega_b = 50~\mathrm{km\,s^{-1}\,kpc^{-1}}$,
as shown in Table~\ref{tabla3}. These orbits correspond to stars with
numbers 6,20,24,25 in this group. The last rows in frames (b) and (d)
of Figure~\ref{fig7}
show the meridional and projected orbits for three of
these stars. Their points in the Bottlinger diagram are four of the
uncircled points in frame (a) of Figure~\ref{fig11}. The resulting
U$^{\prime}$,V$^{\prime}$ velocities for these four non-trapped orbits
are shown in Figure~\ref{fig15}. Each present point is marked with a
colored big circled square, and some lines with the same color define
other velocities in projected positions within the solar vicinity.
These lines lie approximately near the red and blue regions produced by
families V and IX in frame (a) of Figure~\ref{fig11}. Thus, it appears
 that these
four non-trapped orbits could be nearly trapped by these families.
However, there are four other uncircled points in that frame (a) of
Figure~\ref{fig11}
which also lie near the red and blue regions but are trapped by other
families in limited intervals of time, as shown in Table~\ref{tabla3}.
Thus, in the considered solar neighborhood several resonant families
coexist, and giving a point in the Bottlinger diagram is not sufficient
to determine a possible trapping family. The spatial position within
the 3D solar neighborhood is also necessary, apart from the open
problem pointed out in subsection \ref{EJpl} of finding general
conditions for a 3D orbit to be trapped by a 2D resonance on the
Galactic plane. The value of $\Omega_b$ is very important in the
resulting trapping family; as can be seen in Figures~\ref{fig1} and
\ref{fig2} a given orbit may or may not be trapped depending on the
value of $\Omega_b$. In conclusion, we are only confident with the
majority of stars in both groups G18-39 and G21-22 which have been
shown to be trapped by families V and IX, taking $\Omega_b$
approximately in the interval
45--60 $\mathrm{km\,s^{-1}\,kpc^{-1}}$; other stars in these groups
could also be nearly trapped by these families or are trapped by other
less important resonant families.

\section{Conclusions}
\label{concl}

The star groups G18-39 and G21-22 pertaining to the Galactic halo,
and at the present time within the solar neighborhood, were originally
identified by \citet{2012RMxAA..48..109S}. The 3D orbits of these two
groups have been computed in a Galactic potential including a Galactic
bar, and have been analyzed in terms of the orbital structure of
resonant orbits on the Galactic plane created by the bar component.
In a previous study \citep{2015MNRAS.451..705M} we have seen how 3D
star orbits can be trapped by 2D resonant families due to the Galactic
bar. In the present analysis we have shown that the majority of stars
in both groups G18-39 and G21-22 are indeed trapped, mainly by the two
resonant families V and IX studied in \citet{2015MNRAS.451..705M},
taking $\Omega_b$ approximately in the interval
45--60 $\mathrm{km\,s^{-1}\,kpc^{-1}}$.

An  explanation
of the kinematics presented by the stars in these groups was given
by \citet{2012RMxAA..48..109S}. They proposed that the observed LSR
U--V velocity field may be related to the results of simulations made
by \citet{2005MNRAS.359...93M} for the accretion of a dwarf galaxy by
our own Galaxy. Those simulations may give a double-peaked U$^{\prime}$
velocity distribution of stripped stars from the dwarf galaxy, as
observed in particular in the group G18-39. The results that we
obtain with our orbital computations give an alternative explanation
of the kinematics in both groups G18-39 and G21-22, which has to do
closely with the structure of resonant families existing in the
non-axisymmetric potential of our Galaxy. We have shown that the
observed LSR U--V velocity field of the stars in these groups can be
naturally explained as a result of their trapping by resonant
families on the Galactic plane generated by the Galactic bar, mainly
families V and IX. This analysis may help to understand the
identification of other known star groups as the possible result of the
interactions produced by resonances on stars close to resonant
families. For the two groups G18-39 and G21-22 we conclude that the
majority of their stars are members of the supergroups of stars in the
Galaxy trapped by the resonant families V and IX.\\

{\bf Acknowledgements}

We wish to thank greatly Benjam\'in Hern\'andez for his help and support from the Universidad Nacional Aut\'onoma de M\'exico, grant DGAPA PAPIIT IN 100918. E.M acknowledge support from
UNAM/PAPIIT grant IN105916 and IN 114114. J.G.F-T is supported by FONDECYT No. 3180210 and the Chilean BASAL Centro de Excelencia en Astrof\'isica y Tecnolog\'ias Afines (CATA) grant PFB-06/2007.

\end{document}